\documentclass[pra,11pt,onecolumn,notitlepage]{revtex4-2}

\usepackage{etoolbox}
\usepackage{supertabular}
\usepackage{booktabs}
\usepackage{caption}
\usepackage{amsmath}
\usepackage{placeins}
\usepackage{mathtools}
\usepackage{multirow}
\usepackage{braket}
\usepackage{bbm, dsfont}
\usepackage[utf8]{inputenc}
\usepackage{graphicx}
\newcommand{\RN}[1]{\uppercase\expandafter{\romannumeral#1}}
\begin{document}
	\author{K. K. Naseeda}
	\affiliation{Department of Physics, Government College Malappuram, Kerala 676509, India}
    
    \author{Jumana Muhammed Abdul Asharaf}
	\affiliation{%
			Department of Physics, Farook College(Autonomous), Calicut, Kerala 673632, India
	}%
	\title{Quantum Walks Assisted Bidirectional Remote State Preparation}
	\author{N. C. Randeep}
	\affiliation{%
			Department of Physics, Government Arts and Science College, Meenchanda, Calicut, Kerala 673018, India\\
            Department of Physics, SARBTM Govt. College, Koyilandy, Kerala 673307, India
	}%
	
	\date{\today}

\begin{abstract}
 We present a scheme for bidirectional remote state preparation (BRSP) using quantum walks on two independent one-dimensional lattices and two independent cycles with two and four vertices, employing nearest-neighbor jumps with coin outcomes. The protocol is implemented in two distinct ways: one  without the involvement of a controller and the other assisted by a controller . In this approach, the quantum walk dynamics generate entanglement during the remote state preparation process, thereby eliminating the need for any pre-shared initial entanglement between the communicating parties. Finally, we show that the bidirectional remote state preparation schemes based on quantum walks on a two-vertex and four-vertex system exhibit consistent behavior under the respective uncontrolled and controlled configurations.
\end{abstract}
\maketitle
\section{Introduction}

Over the past three decades, quantum communication techniques have advanced significantly. In this development, quantum teleportation \cite{BEN1993,BOU1997,RIG2005,SEB2023,RAN2024}, quantum key distribution \cite{EKE1991,PIR2020,GIS2002}, dense coding \cite{BEN1992,HAR2004,SHA2012}, remote state preparation \cite{LO2000,BEN2001,PET2005,DEV2001,BER2003}, quantum secret sharing \cite{HIL1999,GOT2000,TIT2001,SIN2005}, and quantum dialogue protocols \cite{NGU2004,MAN2005,GAO2010} played a major role. Essentially, all these protocols involve the transmission of quantum information using quantum principles such as superposition and entanglement. Preparing preshared entangled states between communicating parties is a crucial and challenging process in all quantum communication protocols.

 Recently, entanglement generation using quantum walks has been proposed by many researchers and shown to be useful for quantum communication \cite{ABA2006,CAR2005}. A quantum walk is the quantum analogue of a classical random walk governed by quantum principles \cite{VEN2012,KEM2003,KAD2021}. Quantum walk techniques have been experimentally realized in various systems such as trapped ions \cite{ZAR2010}, photonic systems \cite{SCH2010}, and NMR systems \cite{RYA2005}, and it has been observed that they can generate quantum entanglement between the walker and the coin as well as between coins. Hence, quantum walk techniques have been employed in several quantum communication protocols, including quantum teleportation and remote state preparation.

Quantum teleportation is a protocol for transmitting an unknown quantum state from the first party to the second party \cite{BEN1993}, whereas in remote state preparation, the first party prepares a known quantum state at the location of the second party. Remote state preparation was first introduced by Hoi-Kwong Lo \cite{LO2000}, and subsequently many different protocols have been developed \cite{BEN2001,PET2005,DEV2001,BER2003,JIA2021}. Although many protocols for quantum teleportation using quantum walk techniques have been developed \cite{WAN2017,LI2019,SHI2022,ZAR2023,SHA2018,CHA2019,KRI2025}, this approach has not been extensively explored in the context of remote state preparation. Nevertheless, Choudhury et al. have proposed a unidirectional protocol for remote state preparation in the presence of a controller \cite{CHO2024}. However, bidirectional remote state preparation protocols have not yet been explored. In this work, we develop a technique for bidirectional remote state preparation using quantum walks, both with and without a controller. We present the proposed protocol in detail and tabulate all possible measurements and their corresponding unitary transformations for bidirectional remote state preparation (BRSP).

 This paper is organized as follows. In Sect. \ref{sec2}, we introduce uncontrolled bidirectional remote state preparation. In this section, we analyze three cases: first, a walker on a line; second, a walker on a two-vertex complete graph; and third, a walker on a four-cycle. Sect. \ref{sec3} presents controlled bidirectional remote state preparation. There also, we analyze three cases: first, a walker on a line; second, a walker on a two-vertex complete graph; and third, a walker on a four-cycle. Finally, we conclude in Sect. \ref{sec4}.

\section{Uncontrolled Bidirectional Remote State Preparation \label{sec2}}
In bidirectional remote state preparation (BRSP), Alice prepares a state known to her at Bob’s location, and Bob simultaneously prepares a state known to him at Alice’s location using quantum mechanical principles. For such a process to occur, the two communicating parties need to share an entangled state between them. In this BRSP protocol, we introduce a technique in which entanglement is generated within the protocol; this is achieved using quantum walks, specifically coined quantum walks.

In a coined quantum walk, there is a walker and a coin that exist in a combined Hilbert space $\mathcal{H}=\mathcal{H}_{p} \otimes \mathcal{H}_{c}$, where $\mathcal{H}_{p}$ is the Hilbert space of the walker on the lattice and $\mathcal{H}_{c}$ is the Hilbert space of the coin. In this process, the walker on the lattice moves one step to the right  (from site $i$ to $i+1$) when the coin outcome is $0$, and moves one step to the left (from site $i$  to $i-1$) when the coin outcome is $1$, as shown in the FIG. \ref{figure1-quantum walk}. The movement of the walker is implemented by the shift operators defined as follows:

 \begin{figure}
    \centering
    \includegraphics[width=0.9\linewidth]{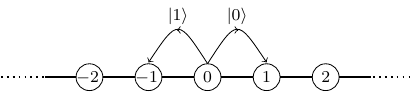}
    \caption{The quantum walk on a one dimensional line. If the coin outcome is $0$, the walker move towards right (from $i$ to $i+1$). If the coin outcome is $1$, the walker moves towards left (from $i$ to $i-1$).}
    \label{figure1-quantum walk}
\end{figure}

\begin{equation}
  \begin{split}
       S_0=&\sum_{i} |i+1\rangle\langle i|\\
       S_0^\dagger=&S_1=\sum_{i}|i\rangle\langle i+1|,
       \label{eqconditional shift}
      \end{split}
 \end{equation}
where $S_0$ and $S_0^{\dagger}=S_1$ are right shift and left shift operators respectively. During the walk, when the coin outcome is $|0\rangle$, the operator $S_0$ acts on the position space and the walker moves one step to the right, whereas when the coin outcome is $|1\rangle$, the operator $S_1$ acts on the position space and the walker moves one step to the left. The coin outcome, followed by the action of the conditional shift operator, executes the quantum walk process. In this section, we use the quantum walk process for uncontrolled bidirectional remote state preparation on a line, a two-vertex complete graph, and a 4-cycle. 
\subsection{Uncontrolled Bidirectional Remote State Preparation Using Quantum Walks on a Line}

Now we discuss the simplest protocol of BRSP, namely BRSP without any controller. In this protocol, there are two parties, Alice and Bob, who are at different spatial locations, and each holds two coins. Alice wants to create a known state $\ket{\phi_1}=a_{0}\ket{0}+a_{1}\ket{1}$ at Bob's place and Bob wants to create a known state $\ket{\phi_2}=b_{0}\ket{0}+b_{1}\ket{1}$ at Alice's place. The coefficients $a_{0}$, $a_{1}$, $b_{0}$ and $b_{1}$ are real and satisfy the conditions $a_{0}^2+a_{1}^2=1$ and $b_{0}^2+b_{1}^2=1$ respectively. Also the state intended for creation is not initially in the physical possession of any of the parties. \par

The present communication task is implemented using quantum walks. For this purpose Alice has three particles: $A_1$, $A_2$ and $A_3$, where the first particle, $A_1$, is in Alice's position space $\mathcal{H}_{P_{A}}$, while the coin spaces $\mathcal{H}_{A_{2}}$ and $\mathcal{H}_{A_{3}}$ of particles $A_2$ and $A_3$, respectively, are the coin spaces held by Alice. Similarly, Bob possesses three particles $B_1$, $B_2$ and $B_3$; particle $B_1$ belongs to Bob’s position space $\mathcal{H}_{P_{B}}$, while particle   $B_2$ and $B_3$ reside in Bob’s coin spaces  $\mathcal{H}_{B_{2}}$ and $\mathcal{H}_{B_{3}}$, respectively.

\begin{figure}
    \centering
    \includegraphics[width=0.9\linewidth]{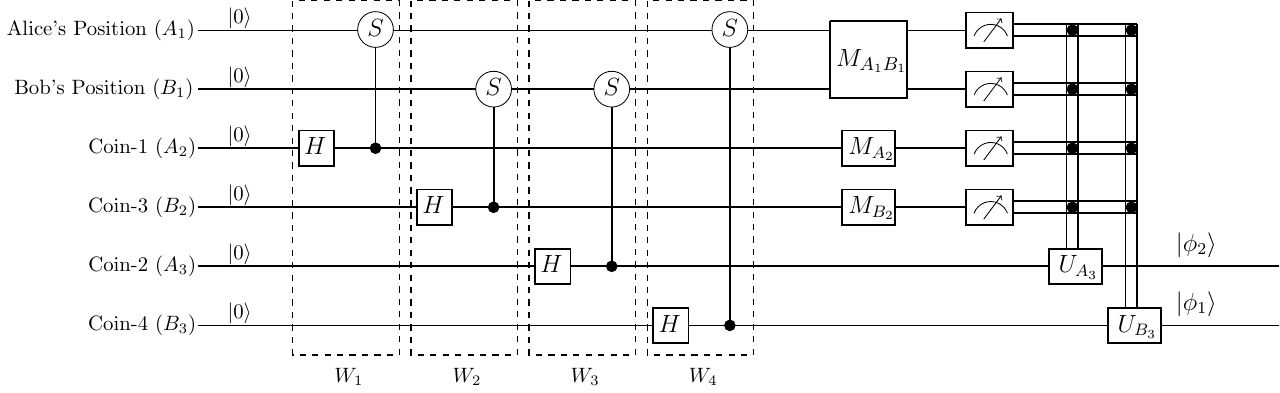}
    \caption{Circuit diagram for uncontrolled bidirectional remote state preparation using quantum walks.}
    \label{figure2-circuit UBRSP1}
\end{figure}

The process of bidirectional remote state preparation involves four distinct walks, namely,
\begingroup
\allowdisplaybreaks
\begin{equation}
\begin{split}
     W_1=E_1(I_{A_1}\otimes I_{B_1}\otimes H_{A_2}\otimes I_{A_3}\otimes I_{B_2}\otimes I_{B_3}) \\
      W_2=E_2(I_{A_1}\otimes I_{B_1}\otimes I_{A_2}\otimes I_{A_3}\otimes H_{B_2}\otimes I_{B_3})\\
      W_3=E_3(I_{A_1}\otimes I_{B_1}\otimes I_{A_2}\otimes H_{A_3}\otimes I_{B_2}\otimes I_{B_3})\\
      W_4=E_4(I_{A_1}\otimes I_{B_1}\otimes I_{A_2}\otimes I_{A_3}\otimes I_{B_2}\otimes H_{B_3}).
\end{split}
\end{equation}
\endgroup
In this case, the walker walks on the line according to the conditional shift operators defined in  Eq.~(\ref{eqconditional shift}), driven by each of the four coins given as follows,

\begingroup
\allowdisplaybreaks
\begin{equation}
    \begin{split}
       E_1=&\sum_{i=0}^1(S_i\otimes I_{B_{1}}\otimes\ket{i}\bra{i}  \otimes I_{A_{3}} \otimes I_{B_{2}} \otimes I_{B_{3}}) \\
       E_{2}=&\sum_{i=0}^1( I_{A_{1}} \otimes S_i \otimes I_{A_{2}}\otimes I_{A_{3}} \otimes\ket{i}\bra{i}\otimes I_{B_{3}}) \\
       E_{3}=&\sum_{i=0}^1( I_{A_1} \otimes S_i \otimes I_{A_2} \otimes \ket{i}\bra{i}\otimes I_{B_2}\otimes I_{B_3} ) \\
       E_4=&\sum_{i=0}^1(S_i\otimes I_{B_1}\otimes I_{A_2}\otimes I_{A_3}\otimes I_{B_2}\otimes \ket{i}\bra{i}). \\
    \end{split}
\end{equation}
\endgroup
Initially, the state of the entire system, consisting of $A_1$, $A_2$, $A_3$, $B_1$, $B_2$ and $B_3$, is given by,
\begin{equation}
    \ket{\psi_0}_{{A_1}{B_1}A_{2}A_{3}B_{2}B_{3}}=\ket{00}\otimes\ket{0}\otimes\ket{0}\otimes\ket{0}\otimes\ket{0}.
    \label{equncontrolled initial}
\end{equation}
After the first step, that is, after the execution of $W_1$, where $H_{A_2}$ is the Hadamard operator acting on coin qubit $A_2$, the initial state given by Eq.~(\ref{equncontrolled initial}) becomes,
\begin{equation}
    \ket{\psi_1}= \frac{1}{\sqrt{2}}(\ket{100000}+\ket{-101000}).
    \label{Eq5}
\end{equation}
Following the second step, that is, the execution of $W_2$, where $H_{B_2}$ is the Hadamard operator acting on coin qubit $B_2$, the state in Eq.~(\ref{Eq5}) becomes,
\begin{equation}
    \ket{\psi_2}= \frac{1}{2}(\ket{110000}+\ket{1-10010}+\ket{-111000}+\ket{-1-11010}).
    \label{Eq6}
\end{equation}
Upon completion of the third step, that is, the execution of $W_3$, where $H_{A_3}$ denote the Hadamard operator acting on coin qubit $A_3$, the state in Eq.~(\ref{Eq6}) becomes,
\begin{equation}
 \begin{split}
    \ket{\psi_3}= &\frac{1}{2\sqrt{2}}(\ket{120000}+\ket{100100}+\ket{100010}+\ket{1-20110}+\\&\ket{-121000}+\ket{-101100}+\ket{-101010}+\ket{-1-21110}).
\end{split}
\label{Eq7}
\end{equation}
Finally, in the fourth step, that is, after the execution of $W_4$, where $H_{B_3}$ is the Hadamard operator acting on coin qubit $B_3$, the state in Eq.~(\ref{Eq7}) becomes,
\begin{equation}
    \begin{split}
        \ket{\psi_4}= &\frac{1}{4}(\ket{220000}+\ket{020001}+\ket{200100}+\ket{000101}+\\&\ket{200010}+\ket{000011}+\ket{2-20110}+\ket{0-20111}+\\&\ket{021000}+\ket{-221001}+\ket{001100}+\ket{-201101}+\\&\ket{001010}+\ket{-201011}+\ket{0-21110}+\ket{-2-21111}).
    \end{split}
    \label{Eq8}
\end{equation}
To proceed further, first Alice and Bob perform measurements on the position spaces $A_1$ and $B_1$ in the bases,
\begin{equation}
    \begin{split}
        M_{A_{1}B_{1}}=&\{\ket{00}, \frac{1}{\sqrt{2}}(\ket{02}+\ket{0-2}), \frac{1}{\sqrt{2}}(\ket{02}-\ket{0-2}), \frac{1}{\sqrt{2}}(\ket{20}+\ket{-20}), \frac{1}{\sqrt{2}}(\ket{20}-\ket{-20}),\\ & \frac{1}{2}(\ket{22}+\ket{2-2}+\ket{-22}+\ket{-2-2}), \frac{1}{2}(\ket{22}-\ket{2-2}+\ket{-22}-\ket{-2-2}), \\&\frac{1}{2}(\ket{22}+\ket{2-2}-\ket{-22}-\ket{-2-2}), \frac{1}{2}(\ket{22}-\ket{2-2}-\ket{-22}+\ket{-2-2} )\}.
    \end{split}
\end{equation}
Subsequently, Alice and Bob perform measurements on the coin spaces $A_2$ and $B_2$ using the bases,
  \begin{equation}
      \begin{split}
          M_{A_{2}}=&\{\ket{\beta_0}=(a_0\ket{0}+a_1\ket{1}),  \ket{\beta_1}=(a_1\ket{0}-a_0\ket{1})\} ~ \text{and}\\
           M_{B_{2}}=&\{\ket{\gamma_0}=(b_0\ket{0}+b_1\ket{1}), \ket{\gamma_1}=(b_1\ket{0}-b_0\ket{1})\},
      \end{split}
  \end{equation}
  respectively.\\
Consider a simple case. Suppose Alice and Bob perform a measurement in the position basis $|00\rangle_{A_{1}B_{1}}$,  then Eq.~(\ref{Eq8}) becomes,
\begin{equation}
    \ket{\psi_5}_{A_2A_3B_2B_3}=\frac{1}{4}(\ket{0101}+\ket{0011}+\ket{1100}+\ket{1010}).
\end{equation}
Then, performing measurements on $A_2$ and $B_2$ in the $\ket{\beta_0}\ket{\gamma_0}$ basis yields,
\begin{equation}
    \ket{\psi_6}_{{A_3}{B_3}}=(b_0\ket{1}+b_1\ket{0})_{A_3}(a_0\ket{1}+a_1\ket{0})_{B_3}
    \label{Eq12}
\end{equation}
with a probability of $\frac{1}{16}$. Finally, to complete BRSP, Alice and Bob perform unitary operations $\sigma^x_{A_3}$ and $\sigma^x_{B_3}$ on the state given in Eq.~(\ref{Eq12}), reducing it to the state, 
\begin{equation}
    \ket{\psi_7}_{{A_3}{B_3}}=(b_0\ket{0}+b_1\ket{1})_{A_3}(a_0\ket{0}+a_1\ket{1})_{B_3}.
\end{equation}
 Thus, Alice and Bob are able to achieve bidirectional remote state preparation between each other. The circuit diagram of the single-qubit BRSP protocol is shown in FIG.\ref{figure2-circuit UBRSP1}. Additionally, all possible measurements and the corresponding unitary transformations for achieving bidirectional remote state preparation are given in Table \ref{tab1}. 
\newpage
\begin{center}
\captionof{table}{Measurements and the corresponding unitary transformations for achieving uncontrolled bidirectional remote state preparation using quantum walks on a line.}
\begin{supertabular}{p{7cm} p{4.5cm} p{4cm}}
    \hline
    Position basis $A_1$ and $B_1$  &   Coin basis of $A_2$ and $B_2$ &  Unitary operations\\
    \hline
       \multirow{4}{*}{$\ket{00}_{A_1B_1}$} &$\ket{\beta_0}_{A_2}\ket{\gamma_0}_{B_2}$&$\sigma^x_{A_3}$$\sigma^x_{B_3}$\\
      \empty &$\ket{\beta_0}_{A_2}\ket{\gamma_1}_{B_2}$&$\sigma^x_{A_3}$$\sigma^z_{A_3}$$\sigma^x_{A_3}$$\sigma^x_{B_3}$\\
        \empty &$\ket{\beta_1}_{A_2}\ket{\gamma_0}_{B_2}$&$\sigma^x_{A_3}$$\sigma^x_{B_3}$$\sigma^z_{B_3}$$\sigma^x_{B_3}$\\
         \empty &$\ket{\beta_1}_{A_2}\ket{\gamma_1}_{B_2}$&$\sigma^x_{A_3}$$\sigma^z_{A_3}$$\sigma^x_{A_3}$$\sigma^x_{B_3}$$\sigma^z_{B_3}$$\sigma^x_{B_3}$\\
         \hline
      \multirow{4}{*}{$\frac{1}{\sqrt{2}}[\ket{02}+\ket{0-2}]_{A_1B_1}$} &$\ket{\beta_0}_{A_2}\ket{\gamma_0}_{B_2}$&$\sigma^x_{B_3}$\\
          \empty&$\ket{\beta_0}_{A_2}\ket{\gamma_1}_{B_2}$&$\sigma^x_{A_3}$$\sigma^z_{A_3}$$\sigma^x_{B_3}$\\
          \empty&$\ket{\beta_1}_{A_2}\ket{\gamma_0}_{B_2}$&$\sigma^x_{B_3}$$\sigma^z_{B_3}$$\sigma^x_{B_3}$\\
          \empty&$\ket{\beta_1}_{A_2}\ket{\gamma_1}_{B_2}$&$\sigma^x_{A_3}$$\sigma^z_{A_3}$$\sigma^x_{B_3}$$\sigma^z_{B_3}$$\sigma^x_{B_3}$\\
          \hline
          \multirow{4}{*}{$ \frac{1}{\sqrt{2}}[\ket{02}-\ket{0-2}]$} &$\ket{\beta_0}_{A_2}\ket{\gamma_0}_{B_2}$&$\sigma^z_{A_3}$$\sigma^x_{B_3}$\\
          \empty&$\ket{\beta_0}_{A_2}\ket{\gamma_1}_{B_2}$&$\sigma^x_{A_3}$$\sigma^x_{B_3}$\\
          \empty&$\ket{\beta_1}_{A_2}\ket{\gamma_0}_{B_2}$&$\sigma^z_{A_3}$$\sigma^x_{B_3}$$\sigma^z_{B_3}$$\sigma^x_{B_3}$\\
          \empty&$\ket{\beta_1}_{A_2}\ket{\gamma_1}_{B_2}$&$\sigma^x_{A_3}$$\sigma^x_{B_3}$$\sigma^z_{B_3}$$\sigma^x_{B_3}$\\
          \hline
          \multirow{4}{*}{$\frac{1}{\sqrt{2}}[\ket{20}+\ket{-20}]_{A_1B_1}$} &$\ket{\beta_0}_{A_2}\ket{\gamma_0}_{B_2}$&$\sigma^x_{A_3}$\\
         \empty&$\ket{\beta_0}_{A_2}\ket{\gamma_1}_{B_2}$&$\sigma^x_{A_3}$ $\sigma^z_{A_3}$$\sigma^x_{A_3}$\\
         \empty&$\ket{\beta_1}_{A_2}\ket{\gamma_0}_{B_2}$&$\sigma^x_{A_3}$ $\sigma^x_{B_3}$$\sigma^z_{B_3}$\\
          \empty&$\ket{\beta_1}_{A_2}\ket{\gamma_1}_{B_2}$&$\sigma^x_{A_3}$ $\sigma^z_{A_3}$ $\sigma^x_{A_3}$$\sigma^x_{B_3}$$\sigma^z_{B_3}$\\
          \hline
          \multirow{4}{*}{$\frac{1}{\sqrt{2}}[\ket{20}-\ket{-20}]_{A_1B_1}$} &$\ket{\beta_0}_{A_2}\ket{\gamma_0}_{B_2}$&$\sigma^x_{A_3}$ $\sigma^z_{B_3}$\\
          \empty&$\ket{\beta_0}_{A_2}\ket{\gamma_1}_{B_2}$&$\sigma^x_{A_3}$ $\sigma^z_{A_3}$$\sigma^x_{A_3}$$\sigma^z_{B_3}$\\
           \empty&$\ket{\beta_1}_{A_2}\ket{\gamma_0}_{B_2}$&$\sigma^x_{A_3}$ $\sigma^x_{B_3}$\\
           \empty&$\ket{\beta_1}_{A_2}\ket{\gamma_1}_{B_2}$&$\sigma^x_{A_3}$ $\sigma^z_{A_3}$$\sigma^x_{A_3}$$\sigma^x_{B_3}$\\
           \hline
           \multirow{4}{*}{$\frac{1}{2}[\ket{22}+\ket{2-2}+\ket{-22}+\ket{-2-2}]_{A_1B_1}$} &$\ket{\beta_0}_{A_2}\ket{\gamma_0}_{B_2}$&I\\
          \empty&$\ket{\beta_0}_{A_2}\ket{\gamma_1}_{B_2}$&$\sigma^x_{A_3}$ $\sigma^z_{A_3}$\\
          \empty&$\ket{\beta_1}_{A_2}\ket{\gamma_0}_{B_2}$&$\sigma^x_{B_3}$ $\sigma^z_{B_3}$\\
          \empty&$\ket{\beta_1}_{A_2}\ket{\gamma_1}_{B_2}$&$\sigma^x_{A_3}$ $\sigma^z_{A_3}$$\sigma^x_{B_3}$$\sigma^z_{B_3}$\\
          \hline
          \multirow{4}{*}{$\frac{1}{2}[\ket{22}-\ket{2-2}+\ket{-22}-\ket{-2-2}]_{A_1B_1}$} &$\ket{\beta_0}_{A_2}\ket{\gamma_0}_{B_2}$&$\sigma^z_{A_3}$\\
          \empty&$\ket{\beta_0}_{A_2}\ket{\gamma_1}_{B_2}$&$\sigma^x_{A_3}$\\
          \empty&$\ket{\beta_1}_{A_2}\ket{\gamma_0}_{B_2}$&$\sigma^z_{A_3}$ $\sigma^x_{B_3}$$\sigma^z_{B_3}$\\
    \empty&$\ket{\beta_1}_{A_2}\ket{\gamma_1}_{B_2}$&$\sigma^x_{A_3}$ $\sigma^x_{B_3}$$\sigma^z_{B_3}$\\
          \hline
          \multirow{4}{*}{$\frac{1}{2}[\ket{22}+\ket{2-2}-\ket{-22}-\ket{-2-2}]_{A_1B_1}$} &$\ket{\beta_0}_{A_2}\ket{\gamma_0}_{B_2}$&$\sigma^z_{B_3}$\\
          \empty&$\ket{\beta_0}_{A_2}\ket{\gamma_1}_{B_2}$&$\sigma^x_{A_3}$ $\sigma^z_{A_3}$$\sigma^z_{B_3}$\\
          \empty&$\ket{\beta_1}_{A_2}\ket{\gamma_0}_{B_2}$&$\sigma^x_{B_3}$\\
          \empty&$\ket{\beta_1}_{A_2}\ket{\gamma_1}_{B_2}$&$\sigma^x_{A_3}\sigma^z_{A_3}\sigma^x_{B_3}$\\
          \hline
          $\frac{1}{2}[\ket{22}-\ket{2-2}-\ket{-22}+\ket{-2-2}]_{A_1B_1}$ &$\ket{\beta_0}_{A_2}\ket{\gamma_0}_{B_2}$&$\sigma^z_{A_3}\sigma^z_{B_3}$\\
          \empty&$\ket{\beta_0}_{A_2}\ket{\gamma_1}_{B_2}$&$\sigma^x_{A_3}$ $\sigma^z_{B_3}$\\
          \empty&$\ket{\beta_1}_{A_2}\ket{\gamma_0}_{B_2}$&$\sigma^z_{A_3}$ $\sigma^x_{B_3}$\\
          \empty&$\ket{\beta_1}_{A_2}\ket{\gamma_1}_{B_2}$&$\sigma^x_{A_3}$ $\sigma^x_{B_3}$\\    
\hline
 \label{tab1}
\end{supertabular}
\end{center}

\subsection{Uncontrolled Bidirectional Remote State Preparation Using Quantum Walks on a Two-Vertex Complete Graph}
\begin{figure}
    \centering
    \includegraphics[width=0.7\linewidth]{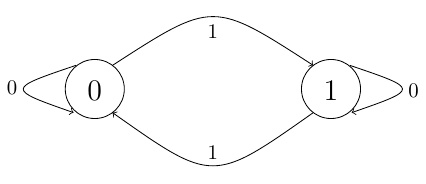}
    \caption{Two vertex complete graph with one coin.}
    \label{figure3-2 vertex circuit}
\end{figure}

For BRSP of single qubit states, we can use quantum walks on a complete graph with two vertices introduced in the reference. For a two-vertex complete graph, there are two vertices, 0 and 1, with two directed edges as shown in FIG. \ref{figure3-2 vertex circuit}. The coin outcome determines the direction of the edge: when the coin outcome is 0, the edge starts from a site and returns to itself, and when the coin outcome is 1, the edge starts from one site and goes to the other. This can be represented by the conditional shift operator defined as follows,
\begin{equation}
    S=(\ket{0}\bra{0}+\ket{1}\bra{1})\otimes \ket{0}\bra{0} +(\ket{1}\bra{0}+\ket{0}\bra{1})\otimes \ket{1}\bra{1}.
    \label{Eq Shift 2 vertex}
\end{equation}
In this conditional shift operator, the walker remains at the same vertex when the coin outcome is 0 and moves to the other vertex when the coin outcome is 1. One thing we must take care of while applying the conditional shift operator is identifying with respect to which coin the conditional shift operation is performed.

As in the previous section, Alice and Bob each have three particles: $A_{1}, A_{2},$ and $A_{3}$ for Alice, and $B_{1}, B_{2},$ and $B_{3}$ for Bob. Here, the first particles of Alice and Bob, $A_1$ and $B_1$, are each defined on independent two-vertex complete graphs. The remaining particles—$A_{2},$ and $A_{3}$ of Alice, and $B_{2},$ and $B_{3}$ of Bob—carry the respective coin states. As earlier, the initial state is given by,
\begin{equation}
\ket{\psi_0}_{{A_1}{B_1}A_{2}A_{3}B_{2}B_{3}}=\ket{00}\otimes\ket{0}\otimes\ket{0}\otimes\ket{0}\otimes\ket{0}.
 \label{Eq15}
\end{equation}
In the first step of the quantum walk, $W_1$, Alice performs a Hadamard operation $H_{A_2}$ on her first coin and then applies the shift operation on Alice’s two-vertex graph, as defined in Eq.~(\ref{Eq Shift 2 vertex}) with respect to Alice’s coin $A_2$. Then. Eq.~(\ref{Eq15}) becomes,
\begin{equation}
    \ket{\psi_1}= \frac{1}{\sqrt{2}}(\ket{000000}+\ket{101000}).
    \label{Eq16}
\end{equation}
After the second step, $W_2$, in which a Hadamard operation $H_{B_2}$ is applied to Bob’s coin $B_{2}$, followed by the shift operation (Eq.(\ref{Eq Shift 2 vertex})) on Bob’s two-vertex graph with respect to Bob’s coin, $B_{2}$, Eq.~(\ref{Eq16}) becomes,
\begin{equation}
     \ket{\psi_2}= \frac{1}{2}(\ket{000000}+\ket{010010}+\ket{101000}+\ket{111010}).
     \label{Eqaa}
\end{equation}
Upon completion of the third step, $W_3$, in which a Hadamard operation $H_{A_3}$ is applied to Alice’s second coin  $A_3$, followed by the shift operation on Bob’s two-vertex graph, as defined in Eq.~(\ref{Eq Shift 2 vertex}), with respect to Alice’s second coin $A_{2}$, Eq.~(\ref{Eqaa}) becomes,
\begin{equation}
 \begin{split}
    \ket{\psi_3}= &\frac{1}{2\sqrt{2}}(\ket{000000}+\ket{010100}+\ket{010010}+\ket{000110}+\\&\ket{101000}+\ket{111100}+\ket{111010}+\ket{101110}).
\end{split}
\label{Eq18}
\end{equation}
Finally, in the fourth walk, $W_4$, Bob performs a Hadamard operation $H_{B_3}$  on his second coin $B_{3}$, followed by the shift operation (Eq.~(\ref{Eq Shift 2 vertex}))  on Alice’s two-vertex graph with respect to Bob’s second coin $B_{3}$. Consequently, Eq.~(\ref{Eq18}) becomes,
\begin{equation}
    \begin{split}
        \ket{\psi_4}= &\frac{1}{4}(\ket{000000}+\ket{100001}+\ket{010100}+\ket{110101}+\\&\ket{010010}+\ket{110011}+\ket{000110}+\ket{100111}+\\&\ket{101000}+\ket{001001}+\ket{111100}+\ket{011101}+\\&\ket{111010}+\ket{011011}+\ket{101110}+\ket{001111}).
    \end{split}
    \label{Eq19}
\end{equation}
Alice and Bob now perform measurements on the position spaces $A_1$ and $B_1$ in the bases,
\begin{equation}
    M_{A_{1}B_{1}}=\{ \ket{\alpha_0}=\ket{00},~ \ket{\alpha_1}=\ket{01},~ \ket{\alpha_2}=\ket{10},~ \ket{\alpha_3}=\ket{11} \}.
\end{equation}
After that, Alice performs measurements on the coin space $A_2$ in the bases,
  \begin{equation}
         M_{A_{2}}=\{ \ket{\beta_0}=(a_0\ket{0}+a_1\ket{1}),~  \ket{\beta_1}=(a_1\ket{0}-a_0\ket{1} \}
  \end{equation}
  and Bob performs measurements on the coin space $B_2$ in the bases,
  \begin{equation}
         M_{B_{2}}=\{ \ket{\gamma_0}=(b_0\ket{0}+b_1\ket{1}),~ \ket{\gamma_1}=(b_1\ket{0}-b_0\ket{1}) \},
  \end{equation}
to achieve remote state preparation. 

If the measurement result in both Alice's and Bob's position basis being $\ket{00}$ , then state in Eq.~(\ref{Eq19}) becomes,
\begin{equation}
    \ket{\psi_5}_{A_2A_3B_2B_3}=\frac{1}{4}(\ket{0000}+\ket{0110}+\ket{1001}+\ket{1111}).
\end{equation}
Finally, on performing measurements in the bases $\ket{\beta_0}\ket{\gamma_0}$ on $A_2$ and $B_2$, yields,
\begin{equation}
    \ket{\psi_7}_{{A_3}{B_3}}=(b_0\ket{0}+b_1\ket{1})_{A_3}(a_0\ket{0}+a_1\ket{1})_{B_3},
\end{equation}
with a probability of $\frac{1}{16}$. Thus, for this particular set of bases for Alice and Bob, BRSP is achieved.
 All remaining possible measurements and the corresponding unitary transformations for achieving bidirectional remote state preparation are given in Table \ref{tab2}.
 \newpage
\begin{center}
\captionof{table}{Measurements and the corresponding unitary transformations for achieving uncontrolled bidirectional remote state preparation using quantum walks on a two-vertex complete graph.}   
\begin{supertabular}{p{5cm} p{4cm} p{4cm}}
\hline
    Position basis $A_1$ and $B_1$  &   Coin basis $A_2$ and $B_2$ &  Unitary operations\\
\hline
\multirow{4}{*}{$\ket{\alpha_0}_{A_1B_1}$} &$\ket{\beta_0}_{A_2}\ket{\gamma_0}_{B_2}$&I\\
          \empty&$\ket{\beta_0}_{A_2}\ket{\gamma_1}_{B_2}$&$\sigma^x_{A_3}$$\sigma^z_{A_3}$\\
          \empty&$\ket{\beta_1}_{A_2}\ket{\gamma_0}_{B_2}$&$\sigma^x_{B_3}$$\sigma^z_{B_3}$\\
          \empty&$\ket{\beta_1}_{A_2}\ket{\gamma_1}_{B_2}$&$\sigma^x_{A_3}$$\sigma^z_{A_3}$$\sigma^x_{B_3}$$\sigma^z_{B_3}$\\
 \hline   \multirow{2}{*}{$\ket{\alpha_1}_{A_1B_1}$} &$\ket{\beta_0}_{A_2}\ket{\gamma_0}_{B_2}$&$\sigma^x_{A_3}$\\
      \empty&$\ket{\beta_0}_{A_2}\ket{\gamma_1}_{B_2}$&$\sigma^x_{A_3}$$\sigma^z_{A_3}$$\sigma^x_{A_3}$\\
      
            \empty&$\ket{\beta_1}_{A_2}\ket{\gamma_0}_{B_2}$&$\sigma^x_{A_3}$$\sigma^x_{B_3}$$\sigma^z_{B_3}$\\
          \empty&$\ket{\beta_1}_{A_2}\ket{\gamma_1}_{B_2}$&$\sigma^x_{A_3}$$\sigma^z_{A_3}$$\sigma^x_{A_3}$$\sigma^x_{B_3}$$\sigma^z_{B_3}$\\
\hline          \multirow{4}{*}{$\ket{\alpha_2}_{A_1B_1}$} &$\ket{\beta_0}_{A_2}\ket{\gamma_0}_{B_2}$&$\sigma^x_{B_3}$\\
          \empty&$\ket{\beta_0}_{A_2}\ket{\gamma_1}_{B_2}$&$\sigma^x_{A_3}$$\sigma^z_{A_3}$$\sigma^x_{B_3}$\\
          \empty&$\ket{\beta_1}_{A_2}\ket{\gamma_0}_{B_2}$&$\sigma^x_{B_3}$$\sigma^z_{B_3}$$\sigma^x_{B_3}$\\
          \empty&$\ket{\beta_1}_{A_2}\ket{\gamma_1}_{B_2}$&$\sigma^x_{A_3}$$\sigma^z_{A_3}$$\sigma^x_{B_3}$$\sigma^z_{B_3}$$\sigma^x_{B_3}$\\
\hline           \multirow{4}{*}{$\ket{\alpha_3}_{A_1B_1}$} &$\ket{\beta_0}_{A_2}\ket{\gamma_0}_{B_2}$&$\sigma^x_{A_3}$$\sigma^x_{B_3}$\\
       \empty &$\ket{\beta_0}_{A_2}\ket{\gamma_1}_{B_2}$&$\sigma^x_{A_3}$$\sigma^z_{A_3}$$\sigma^x_{A_3}$$\sigma^x_{B_3}$\\
        \empty &$\ket{\beta_1}_{A_2}\ket{\gamma_0}_{B_2}$&$\sigma^x_{A_3}$$\sigma^x_{B_3}$$\sigma^z_{B_3}$$\sigma^x_{B_3}$\\
         \empty &$\ket{\beta_1}_{A_2}\ket{\gamma_1}_{B_2}$&$\sigma^x_{A_3}$$\sigma^z_{A_3}$$\sigma^x_{A_3}$$\sigma^x_{B_3}$$\sigma^z_{B_3}$$\sigma^x_{B_3}$\\
\hline
 \label{tab2}
\end{supertabular}
\end{center}
\FloatBarrier
\subsection{Uncontrolled Bidirectional Remote State Preparation Using Quantum Walks on 4-Cycle}

\begin{figure} 
    \centering
    \includegraphics[width=0.6\linewidth]{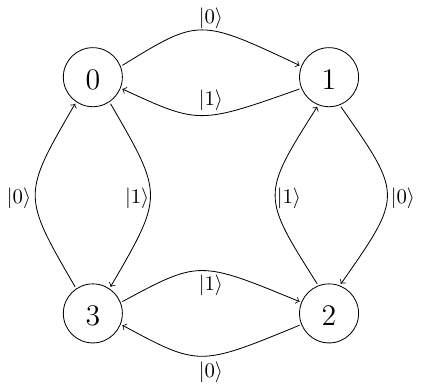}
    \caption{A cycle with four vertices (4-cycle).}
    \label{fig:fig2}
\end{figure}

In this BRSP scheme, the position spaces of Alice and Bob are two independent 4-cycle, with vertices labeled $0$, $1$, $2$ and $3$ in each, as shown in FIG. \ref{fig:fig2}. As in the previous section, Alice has three particles: $A_1$, $A_2$ and $A_3$, where the state of $A_1$ represents Alice’s position space (4-cycle), and the states of $A_2$ and $A_3$ are Alice’s first and second coin spaces, respectively. Similarly, $B_1$ represents Bob’s position space (4-cycle), and the states of $B_2$ and $B_3$ are Bob’s first and second coin spaces, respectively. The conditional shift operator in the 4-cycle BRSP is defined as,

 \begin{equation}
  \begin{split}
       S = \sum_{i=0}^{2} \ket{i+1}\bra{i} \otimes \ket{0}\bra{0}
+ \sum_{i=0}^{3} \ket{i-1}\bra{i} \otimes \ket{1}\bra{1}
+ \ket{0}\bra{3} \otimes \ket{0}\bra{0}
+ \ket{3}\bra{0} \otimes \ket{1}\bra{1}.
\label{eq25}
       \end{split}
 \end{equation}
This conditional shift is equivalent to the conditional shift operator defined in Eq.~(\ref{eqconditional shift});   however, in the present case only four vertices are involved, and their periodic connectivity (as shown in FIG \ref{fig:fig2}), introduces periodicity into the conditional shift operator. Also, this conditional shift operates with respect to a specified coin, which must be mentioned during the walk. 

Initially, the position spaces of $A_1$ and $B_1$ are initialized to $\ket{0}$, and the coin spaces of $A_2$,$A_3$,$B_2$ and $B_3$ are initialized to $\ket{0}$. Thus, the total initial state, in a convenient order is

\begin{equation}
    \ket{\psi_{0}}_{A_{1}B_{1}A_{2}A_{3}B_{2}B_{3}}=\ket{00}\otimes\ket{0}\otimes\ket{0}\otimes\ket{0}\otimes\ket{0}.
    \label{eq}
 \end{equation}
After the first step $W_1$, that is, after the execution of $H_{A_2}$ followed by the conditional shift operation on the Alice's 4-cycle given in Eq.~(\ref{eq25}) with respect to Alice’s first coin  $A_{2}$, the above Eq.~(\ref{eq}) becomes,
\begin{equation}
    \ket{\psi_1}= \frac{1}{\sqrt{2}}(\ket{100000}+\ket{301000}).
    \label{Eq27}
\end{equation}
Following the second step $W_2$, in which $H_{B_2}$ is applied followed by the conditional shift operation on the Bob's 4-cycle given in Eq.~(\ref{eq25}) with respect to Bob’s first coin  $B_2$, the state in Eq.~ (\ref{Eq27}) becomes,
\begin{equation}
     \ket{\psi_2}= \frac{1}{2}(\ket{110000}+\ket{130010}+\ket{311000}+\ket{331010}).
     \label{equ}
\end{equation}
Then we proceed to the third step $W_3$, that is, after the execution of $H_{A_3}$ followed by the conditional shift operation on the Bob's 4-cycle given in Eq.~(\ref{eq25}) with respect to Alice’s second coin $A_3$, the state in Eq.~(\ref{equ}) becomes,
\begin{equation}
 \begin{split}
    \ket{\psi_3}= &\frac{1}{2\sqrt{2}}(\ket{120000}+\ket{100100}+\ket{100010}+\ket{120110}+\\&\ket{321000}+\ket{301100}+\ket{301010}+\ket{321110}).
\end{split}
\label{equa}
\end{equation}
Finally, in the fourth step $W_4$, Bob performs the operation $H_{B_3}$ on his second coin, followed by the conditional shift operation on the Alice's 4-cycle given in Eq.~(\ref{eq25})  with respect to Bob’s second coin $B_3$, the state in Eq.~(\ref{equa}) becomes,
\begin{equation}
    \begin{split}
        \ket{\psi_4}= &\frac{1}{4}(\ket{220000}+\ket{020001}+\ket{200100}+\ket{000101}+\\&\ket{200010}+\ket{000011}+\ket{220110}+\ket{020111}+\\&\ket{021000}+\ket{221001}+\ket{001100}+\ket{201101}+\\&\ket{001010}+\ket{201011}+\ket{021110}+\ket{221111}).
    \end{split}
    \label{Equa}
\end{equation}
Now Alice and Bob perform measurements on the position spaces $A_1$ and $B_1$ in the bases:
\begin{equation}
    M_{A_{1}B_{1}}= \{ (\ket{\tilde{\alpha_0}}=\ket{22},~\ket{\tilde{\alpha_1}}=\ket{20},~\ket{\tilde{\alpha_2}}=\ket{02},~ \ket{\tilde{\alpha_3}}=\ket{00}) \}.
    \label{Eq 4cycle positon basis}
\end{equation}
After that, Alice performs measurements on the coin space $A_2$ in the bases,
  \begin{equation}
         M_{A_{2}}=\{ \ket{\beta_0}=(a_0\ket{0}+a_1\ket{1}),~  \ket{\beta_1}=(a_1\ket{0}-a_0\ket{1}) \}
  \end{equation}
  and Bob performs measurements on the coin space $B_2$ in the bases,
  \begin{equation}
         M_{B_{2}}=\{ \ket{\gamma_0}=(b_0\ket{0}+b_1\ket{1}),~ \ket{\gamma_1}=(b_1\ket{0}-b_0\ket{1}) \},
  \end{equation}
to achieve BRSP. 

 If the measurement result in both Alice's and Bob's position basis being $\ket{22}$, then state in Eq.~(\ref{Equa}) becomes,
\begin{equation}
    \ket{\psi_5}_{A_2A_3B_2B_3}=\frac{1}{4}(\ket{0000}+\ket{0110}+\ket{1001}+\ket{1111}).
\end{equation}
Finally, on performing measurements in the bases $\ket{\beta_0}\ket{\gamma_0}$ on $A_2$ and $B_2$, yields,
\begin{equation}
    \ket{\psi_7}_{{A_3}{B_3}}=(b_0\ket{0}+b_1\ket{1})_{A_3}(a_0\ket{0}+a_1\ket{1})_{B_3},
\end{equation}
with a probability of $\frac{1}{16}$. Thus, for this particular set of bases for Alice and Bob, BRSP is achieved. All remaining possible measurements and the corresponding unitary transformations for achieving bidirectional remote state preparation are equivalent to the two-vertex case given in Table \ref{tab2}. More specifically, to obtain the exact table in this case, one only needs to replace the position basis $|\alpha_{i}\rangle$ introduced in Table \ref{tab2} with  $|\tilde{\alpha_{i}}\rangle$ defined in Eq.~(\ref{Eq 4cycle positon basis}).

\section{Controlled Bidirectional Remote State Preparation Using Quantum Walks \label{sec3}}

In the preceding section, we analyzed the BRSP protocol without any controller. In this section, we analyze the BRSP protocol with the presence of a controller, Charlie. In this protocol, as in the previous section, Alice and Bob each hold three particles, $A_1, A_2, A_3$ and $B_1, B_2, B_3$, respectively, whose roles are the same as before. In addition, Charlie holds two particles, $C_1$ and $C_2$, which control the remote state preparation of the two parties. Unlike uncontrolled BRSP, Alice and Bob cannot remotely prepare a state without Charlie’s permission. This increases the security of the protocol and makes it an authorized BRSP protocol. As in the previous section, here also we introduce a controlled BRSP protocol on a line, a two-vertex graph, and a 4-cycle.

\subsection{Controlled Bidirectional Remote State Preparation Using Quantum Walks on a Line}

In this section, we discuss the controlled BRSP of a single qubit using a controller on a line. The situation is such that there are two parties, Alice and Bob, located at different spatial positions and we have a controller, Charlie, who holds two coins for their actions.  Alice wants to create a known state $\ket{\phi_1}=a_0\ket{0}+a_1\ket{1}$ at Bob’s place, and  Bob wants to create a known state $\ket{\phi_2}=b_0\ket{0}+b_1\ket{1}$ at Alice’s place; this is the task of the protocol. In this case, the conditional shift operator is identical to the one defined in Eq.~(\ref{eqconditional shift}) in Sect. I.

Initially,  the position spaces of $A_1$ and $B_1$ initialized to $\ket{0}$, and coin spaces of $A_2$,$A_3$,$B_2$,$B_3$, $C_1$ and $C_2$ are initialized to $\ket{0}$. Thus, the total initial state in the convenient order is,

\begin{equation}
    \ket{\psi_{0}}_{A_{1}B_{1}A_{2}A_{3}B_{2}B_{3}C_{1}C_{2}}=\ket{00}\otimes\ket{0}\otimes\ket{0}\otimes\ket{0}\otimes\ket{0}\otimes\ket{0}\otimes\ket{0}.
    \label{Eq17}
 \end{equation}
The controlled BRSP protocol involves six distinct walks, which are described as follows:

\begin{equation}
\begin{split}
      W_1=E_1(I_{A_1}\otimes I_{B_1}\otimes H_{A_2}\otimes I_{A_3}\otimes I_{B_2}\otimes I_{B_3}\otimes I_{C_1}\otimes I_{C_2}) \\
   W_2=E_2(I_{A_1}\otimes I_{B_1}\otimes I_{A_2}\otimes I_{A_3}\otimes H_{B_2}\otimes I_{B_3}\otimes I_{C_1}\otimes I_{C_2}) \\
  W_3=E_3(I_{A_1}\otimes I_{B_1}\otimes I_{A_2}\otimes H_{A_3}\otimes I_{B_2}\otimes I_{B_3}\otimes I_{C_1}\otimes I_{C_2}) \\
   W_4=E_4(I_{A_1}\otimes I_{B_1}\otimes I_{A_2}\otimes I_{A_3}\otimes I_{B_2}\otimes H_{B_3}\otimes I_{C_1}\otimes I_{C_2}) \\
     W_5=E_3(I_{A_1}\otimes I_{B_1}\otimes I_{A_2}\otimes I_{A_3}\otimes I_{B_2}\otimes I_{B_3}\otimes H_{C_1}\otimes I_{C_2}) \\
   W_4=E_4(I_{A_1}\otimes I_{B_1}\otimes I_{A_2}\otimes I_{A_3}\otimes I_{B_2}\otimes I_{B_3}\otimes I_{C_1}\otimes H_{C_2}), \\
\end{split}
\end{equation}
where,

\begin{equation}
 \begin{split}
E_1=&\sum_{i=0}^1(S_i\otimes I_{B_{1}}\otimes\ket{i}\bra{i}  \otimes I_{A_{3}} \otimes I_{B_{2}} \otimes I_{B_{3}}\otimes I_{C_1}\otimes I_{C_2}) \\
E_{2}=&\sum_{i=0}^1( I_{A_{1}} \otimes S_i \otimes I_{A_{2}}\otimes I_{A_{3}} \otimes\ket{i}\bra{i}\otimes I_{B_{3}}\otimes I_{C_1}\otimes I_{C_2}) \\
E_{3}=&\sum_{i=0}^1( I_{A_1} \otimes S_i \otimes I_{A_2} \otimes \ket{i}\bra{i}\otimes I_{B_2}\otimes I_{B_3}\otimes I_{C_1}\otimes I_{C_2} ) \\
E_4=&\sum_{i=0}^1(S_i\otimes I_{B_1}\otimes I_{A_2}\otimes I_{A_3}\otimes I_{B_2}\otimes \ket{i}\bra{i} \otimes I_{C_1}\otimes I_{C_2})\\
E_5=&\sum_{i=0}^1(S_i\otimes I_{B_1}\otimes I_{A_2}\otimes I_{A_3}\otimes I_{B_2}\otimes I_{B_3}\otimes \ket{i}\bra{i}\otimes I_{C_2} )\\
E_{6}=&\sum_{i=0}^1( I_{A_1} \otimes S_i \otimes I_{A_2} \otimes I_{A_3} \otimes I_{B_2}\otimes I_{B_3}\otimes I_{C_1}\otimes \ket{i}\bra{i} ). 
\end{split}
\end{equation}

Now, we analyze the protocol in detail by first applying the initial step of the quantum walk, $W_1$, to Eq.~(\ref{Eq17})  we get,

  \begin{equation}
    \ket{\psi_{1}}=\frac{1}{\sqrt{2}}(\ket{10000000}+\ket{-10100000}).
    \end{equation}

After applying the second step, $W_2$, we obtain the state,
 
 \begin{equation}
    \ket{\psi_{2}}=\frac{1}{2}(\ket{11000000}+\ket{1-1001000}+\ket{-11100000}+\ket{-1-1101000}).
    \end{equation}

Next, applying the third step of the quantum walk, $W_3$, transforms the above state into,
\begin{equation}
\begin{split}
\ket{\psi_{3}}=&\frac{1}{2\sqrt{2}}(\ket{12000000}+\ket{10010000}+\ket{10001000}+\ket{1-2011000} \\&+\ket{-12100000}+\ket{-10110000}+\ket{-10101000}+\ket{-1-2111000}).    
\end{split}
\end{equation}

Then, the action of $W_4$ yields,

\begin{equation}
\begin{split}
\ket{\psi_{4}}=&\frac{1}{4}(\ket{22000000}+\ket{02000100}+\ket{20010000}+\ket{00010100}\\
&+\ket{20001000}+\ket{00001100}+\ket{2-2011000}+\ket{0-2011100}\\ &+\ket{02100000}+\ket{-22100100}+\ket{00110000}+\ket{-20110100}\\
&+\ket{-00101000}+\ket{-20101100}+\ket{0-2111000}+\ket{-2-2111100}).   
\end{split}
\end{equation}

Following this, the action of $W_5$ gives,

\begin{equation}
    \begin{split}
       \ket{\psi_{5}}=&\frac{1}{4\sqrt{2}}(\ket{32000000}+\ket{12000010}
    +\ket{12000100}+\ket{-12000110}\\
       &+\ket{30010000}+\ket{10010010}+\ket{10010100}+\ket{-10010110}+\\
       &+\ket{30001000}+\ket{10001010}+\ket{10001100}+\ket{-10001110}+\\
       &+\ket{3-2011000}+\ket{1-2011010}+\ket{1-2011100}+\ket{-1-2011110}\\
       &+\ket{12100000}+\ket{-12100010}+\ket{-12100100}+\ket{-32100110}\\
       &+\ket{10110000}+\ket{-10110010}+\ket{-10110100}+\ket{-30110110}+\\
       &+\ket{10101000}+\ket{-10101010}+\ket{-10101100}+\ket{-30101110}\\
       &+\ket{1-2111000}+\ket{-1-2111010}+\ket{-1-2111100}+\ket{-3-2111110}).
    \end{split}
\end{equation}

Finally, after applying $W_6$, we obtain the final state,
\begingroup
\allowdisplaybreaks
\begin{equation}
    \begin{split}
       \ket{\psi_{6}}=&\frac{1}{8}(\ket{33000000}+\ket{31000001}+\ket{13000010}+\ket{11000011}\\
       &+\ket{13000100}+\ket{11000101}+\ket{-13000110}+\ket{-11000111}\\
       &+\ket{31010000}+\ket{3-1010001}+\ket{11010010}+\ket{1-1010011}\\
       &+\ket{11010100}+\ket{1-1010101}+\ket{-11010110}+\ket{-1-1010111}\\
       &+\ket{31001000}+\ket{3-1001001}+\ket{11001010}+\ket{1-1001011}\\
       &+\ket{11001100}+\ket{1-1001101}+\ket{-11001110}+\ket{-1-1001111}\\
       &+\ket{3-1011000}+\ket{3-3011001}+\ket{1-1011010}+\ket{1-3011011}\\
       &+\ket{1-1011100}+\ket{1-3011101}+\ket{-1-1011110}+\ket{-1-3011111}\\
       &+\ket{13100000}+\ket{11100001}+\ket{-13100010}+\ket{-11100011}\\
       &+\ket{-13100100}+\ket{-11100101}+\ket{-33100110}+\ket{-31100111}\\
       &+\ket{11110000}+\ket{1-1110001}+\ket{-11110010}+\ket{-1-1110011}\\
       &+\ket{-11110100}+\ket{-1-1110101}+\ket{-31110110}+\ket{-3-1110111}\\
       &+\ket{11101000}+\ket{1-1101001}+\ket{-11101010}+\ket{-1-1101011}\\
       &+\ket{-11101100}+\ket{-1-1101101}+\ket{-31101110}+\ket{-3-1101111}\\
       &+\ket{1-1111000}+\ket{1-3111001}+\ket{-1-1111010}+\ket{-1-3111011}\\
       &+\ket{-1-1111100}+\ket{-1-3111101}+\ket{-3-1111110}+\ket{-3-3111111}).\\
       \label{Eq25}
    \end{split}
\end{equation}
\endgroup
To complete the protocol, we perform measurements on all position states and selected components of the coin states. If the measurement results in both Alice’s and Bob’s position bases being $\frac{1}{2}(\ket{33}+\ket{3-1}+\ket{-13}+\ket{-1-1})$, then Eq.~(\ref{Eq25}) becomes,

 \begin{equation}
 \begin{split}
     \ket{\psi_{7}}=&\frac{1}{16}(\ket{000000}+\ket{000110}+\ket{010000}+\ket{010011}\\
     &+\ket{001001}+\ket{001111}+\ket{011000}+\ket{011110}\\
     &+\ket{100010}+\ket{100100}+\ket{110011}+\ket{110101}\\
     &+\ket{101011}+\ket{101101}+\ket{111010}+\ket{111100}).
     \end{split}
 \end{equation}
As in the previous section, here also Alice ($A_{2}$) and Bob ($B_{2}$) can perform measurements on the following set of coin bases:
 \begin{equation}
      M_{A_{2}}= \{\ket{\beta_{0}}_{A_2}=(a_{0}\ket{0}+a_{1}\ket{1}),~               \ket{\beta_{1}}_{A_2}=(a_{1}\ket{0}-a_{0}\ket{1}) \}
      \label{eq46}
    \end{equation}
and
 \begin{equation}
       M_{B_{2}}= \{ \ket{\gamma_{0}}_{B_2}=(b_{0}\ket{0}+b_{1}\ket{1}),~               \ket{\gamma_{1}}_{B_2}=(b_{1}\ket{0}-b_{0}\ket{1}) \}.
       \label{eq47}
 \end{equation}
Suppose, Alice and Bob perform measurements on coins $A_2$ and $B_2$, respectively, in the bases $\ket{\beta_0}\ket{\gamma_0}$; this yields,

 \begin{equation}
 \begin{split}
     \ket{\psi_{8}}=&a_{0}b_{0}\ket{0000}+a_{0}b_{0}\ket{0110}+a_{0}b_{0}\ket{1000}+a_{0}b_{0}\ket{1011}\\
     &+a_{0}b_{1}\ket{0001}+a_{0}b_{1}\ket{0111}+a_{0}b_{1}\ket{1000}+a_{0}b_{1}\ket{1110}\\
     &+a_{1}b_{0}\ket{0010}+a_{1}b_{0}\ket{0100}+a_{1}b_{0}\ket{1011}+a_{1}b_{0}\ket{1101}\\
     &+a_{1}b_{1}\ket{0011}+a_{1}b_{1}\ket{0101}+a_{1}b_{1}\ket{1010}+a_{1}b_{1}\ket{1100}.
     \end{split}
 \end{equation}
Finally, Charlie performs measurements on particles $C_1$ and $C_2$ in the $\ket{00}$ basis, yielding,

\begin{equation}
 \begin{split}
     \ket{\psi_{9}}=(b_{0}\ket{0}+b_{1}\ket{1})_{A_{3}}(a_{0}\ket{0}+a_{1}\ket{1})_{B_{3}}
     \end{split}
 \end{equation}

with a probability of $\frac{1}{256}$. That is, Alice is able to create a known state at Bob’s location, and Bob is able to create a known state at Alice’s location simultaneously. Here, the measurement outcomes of Charlie and the associated classical communication play a crucial role in the protocol; without the controller Charlie’s permission, the protocol cannot be completed. Thus, the presence of a controller makes it an authorized protocol. Instead of measuring in the $\ket{00}$ basis, $C_1$ and $C_2$ can perform measurements in the $\ket{01}$, $\ket{10}$ and $\ket{11}$ bases. All these meausurements results are shown in Appendix 1.
Table. \ref{table2}. In addition, Alice and Bob have to measure the remaining position bases to complete the protocol. The suitable position basis sets of Alice and Bob, $A_{1}$ and $B_{1}$, respectively, for the controlled BRSP protocol using quantum walks on a line are,
\begingroup
\allowdisplaybreaks
\begin{equation}
    \begin{split}
       &M_{A_{1}B_{1}}=\\\{ &\ket{\alpha_{0}}=\frac{1}{2}(\ket{33}+\ket{3-1}+\ket{-13}+\ket{-1-1}),~~
        \ket{\alpha_{1}}=\frac{1}{2}(\ket{33}-\ket{3-1}+\ket{-13}-\ket{-1-1})\\
        &\ket{\alpha_{2}}=\frac{1}{2}(\ket{33}+\ket{3-1}-\ket{-13}-\ket{-1-1}),~~
        \ket{\alpha_{3}}=\frac{1}{2}(\ket{33}-\ket{3-1}-\ket{-13}+\ket{-1-1})\\
        &\ket{\alpha_{4}}=\frac{1}{2}(\ket{31}+\ket{3-3}+\ket{-11}+\ket{-1-3}),~~
        \ket{\alpha_{5}}=\frac{1}{2}(\ket{31}-\ket{3-3}+\ket{-11}-\ket{-1-3})\\
        &\ket{\alpha_{6}}=\frac{1}{2}(\ket{31}+\ket{3-3}-\ket{-11}-\ket{-1-3}),~~
        \ket{\alpha_{7}}=\frac{1}{2}(\ket{31}-\ket{3-3}-\ket{-11}+\ket{-1-3})\\
        &\ket{\alpha_{8}}=\frac{1}{2}(\ket{13}+\ket{1-1}+\ket{-33}+\ket{-3-1}),~~
        \ket{\alpha_{9}}=\frac{1}{2}(\ket{13}-\ket{1-1}+\ket{-33}-\ket{-3-1})\\
        &\ket{\alpha_{10}}=\frac{1}{2}(\ket{13}+\ket{1-1}-\ket{-33}-\ket{-3-1}),~~
        \ket{\alpha_{11}}=\frac{1}{2}(\ket{13}-\ket{1-1}-\ket{-33}+\ket{-3-1})\\
        &\ket{\alpha_{12}}=\frac{1}{2}(\ket{11}+\ket{1-3}+\ket{-31}+\ket{-3-3}),~~
        \ket{\alpha_{13}}=\frac{1}{2}(\ket{11}-\ket{1-3}+\ket{-31}-\ket{-3-3})\\
        &\ket{\alpha_{14}}=\frac{1}{2}(\ket{11}+\ket{1-3}-\ket{-31}-\ket{-3-3}),~
        \ket{\alpha_{15}}=\frac{1}{2}(\ket{11}-\ket{1-3}-\ket{-31}+\ket{-3-3})\}.\\
        \end{split}
\end{equation}
\endgroup

\begin{figure}
    \centering
    \includegraphics[width=0.9\linewidth]{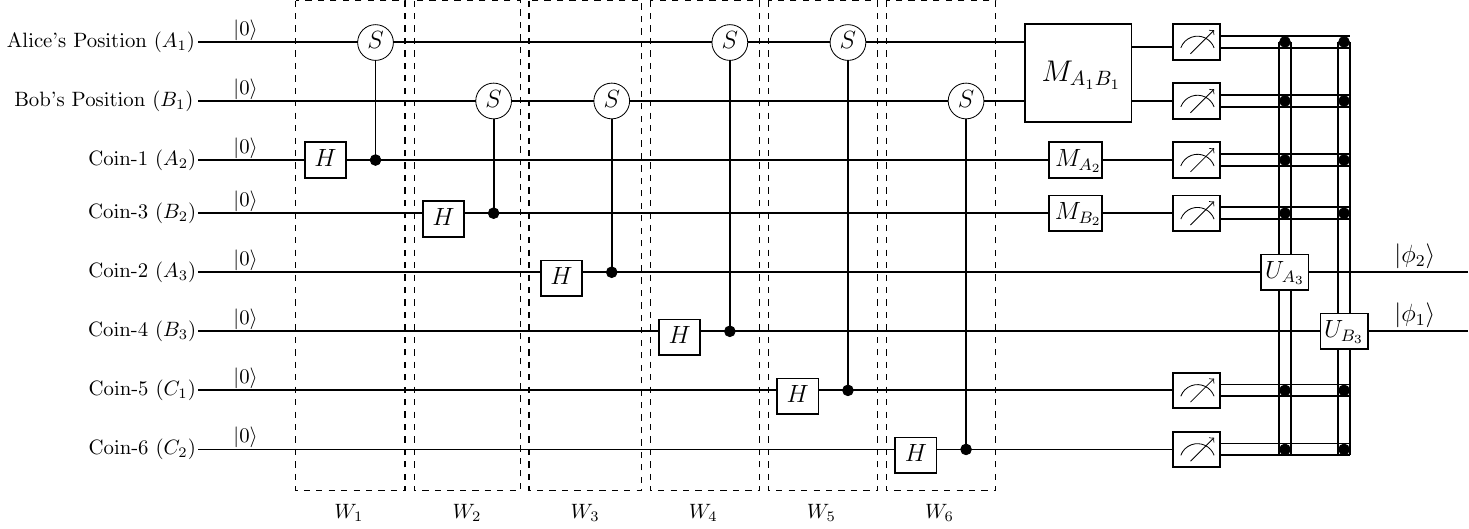}
    \caption{Circuit diagram for controlled bidirectional remote state preparation using quantum walks.}
    \label{figure4-circuit CBRSP2}
\end{figure}

 Alice and Bob perform measurements in the above set of bases $M_{A_{1},B_{1}}$, followed by measurements of the respective coin bases by Alice, Bob, and Charlie, as discussed earlier. The corresponding unitary operations that complete the protocol are shown in Appendix A, Table \ref{table2}. The circuit diagram of this controlled BRSP protocol is shown in FIG. \ref{figure4-circuit CBRSP2}.

\subsection{Controlled Bidirectional Remote State Preparation Using Quantum Walks on a Two-Vertex Complete Graph}

In this controlled BRSP scheme, the position spaces (as shown in FIG.\ref{figure3-2 vertex circuit}) and the coin spaces of Alice and Bob are the same as those in the two independent 2-vertex case discussed in the uncontrolled protocol. In this controlled protocol, in addition, there are two coin spaces $C_1$ and $C_2$ for the controller.

Initially, the position spaces $A_1$ and $B_1$ are initialized to $\ket{0}$, and the coin spaces  $A_2$,$A_3$,$B_2$,$B_3$, $C_1$ and $C_2$ are initialized to $\ket{0}$. Thus, the total initial state, in a convenient order, is,

\begin{equation}
    \ket{\psi_{0}}_{A_{1}B_{1}A_{2}A_{3}B_{2}B_{3}C_{1}C_{2}}=\ket{00}\otimes\ket{0}\otimes\ket{0}\otimes\ket{0}\otimes\ket{0}\otimes\ket{0}\otimes\ket{0}.
    \label{eqn62}
 \end{equation}
 The conditional shift operator and the first four walks are the same as in the uncontrolled 2-vertex BRSP protocol, and in addition, there are two more walks controlled by Charlie. After the first step of the quantum walk, $W_1$, Eq.~(\ref{eqn62}) becomes,
  \begin{equation}
    \ket{\psi_{1}}=\frac{1}{\sqrt{2}}(\ket{10000000}+\ket{10100000}).
    \label{Eq2}
 \end{equation}

Following this, upon applying the second step of quantum walk, $W_2$, we obtain the state,
 
 \begin{equation}
    \ket{\psi_{2}}=\frac{1}{2}(\ket{00000000}+\ket{01001000}+\ket{10100000}+\ket{11101000}).
    \label{Eq2a}
 \end{equation}

Next, applying the third step of the quantum walk, $W_{3}$, transforms the above state to,

\begin{equation}
\begin{split}
\ket{\psi_{3}}=&\frac{1}{2}(\ket{00000000}+\ket{01010000}+\ket{01001000}+\ket{00011000} \\&+\ket{10100000}+\ket{11110000}+\ket{11101000}+\ket{10101000}).   
\end{split}
\end{equation}

Then, the action of $W_4$ yields,

\begin{equation}
\begin{split}
\ket{\psi_{4}}=&\frac{1}{4}(\ket{00000000}+\ket{10000100}+\ket{01010000}+\ket{11010100}\\
&+\ket{01001000}+\ket{11001100}+\ket{00011000}+\ket{10011100}\\ &+\ket{10100000}+\ket{00100100}+\ket{11110000}+\ket{01110100}\\
&+\ket{11101000}+\ket{01101100}+\ket{10111000}+\ket{00111100}).    
\end{split}
\end{equation}

Following this, after the fifth step of the walk, $W_5$, in which a controlled shift operator acts on Alice’s position with respect to Charlie’s first coin $C_{1}$, the above state becomes,

\begin{equation}
    \begin{split}
       \ket{\psi_{5}}=&\frac{1}{4\sqrt{2}}(\ket{00000000}+\ket{10000010}
    +\ket{10000100}+\ket{00000110}\\
       &+\ket{00010000}+\ket{10010010}+\ket{10010100}+\ket{00010110}+\\
       &+\ket{00001000}+\ket{10001010}+\ket{10001100}+\ket{00001110}+\\
       &+\ket{01011000}+\ket{11011010}+\ket{11011100}+\ket{01011110}\\
       &+\ket{10100000}+\ket{01100010}+\ket{00100100}+\ket{10100110}\\
       &+\ket{10110000}+\ket{00110010}+\ket{00110100}+\ket{10110110}+\\
       &+\ket{10101000}+\ket{00101010}+\ket{00101100}+\ket{10101110}\\
       &+\ket{11111000}+\ket{01111010}+\ket{01111100}+\ket{11111110}).
    \end{split}
\end{equation}

Finally, in the sixth step, $W_6$, a controlled shift operator acts on Bob’s position with respect to Charlie’s second coin $C_{2}$, and the above state becomes,
\begingroup
\allowdisplaybreaks
\begin{equation}
    \begin{split}
       \ket{\psi_{6}}=&\frac{1}{8}(\ket{00000000}+\ket{01000001}+\ket{10000010}+\ket{11000011}\\
       &+\ket{10000100}+\ket{11000101}+\ket{00000110}+\ket{01000111}\\
       &+\ket{01010000}+\ket{00010001}+\ket{11010010}+\ket{10010011}\\
       &+\ket{11010100}+\ket{10010101}+\ket{01010110}+\ket{00010111}\\
       &+\ket{01001000}+\ket{00001001}+\ket{11001010}+\ket{10001011}\\
       &+\ket{11001100}+\ket{10001101}+\ket{01001110}+\ket{00001111}\\
       &+\ket{00011000}+\ket{01011001}+\ket{10011010}+\ket{11011011}\\
       &+\ket{10011100}+\ket{11011101}+\ket{00011110}+\ket{01011111}\\
       &+\ket{10100000}+\ket{11100001}+\ket{00100010}+\ket{01100011}\\
       &+\ket{00100100}+\ket{01100101}+\ket{10100110}+\ket{11100111}\\
       &+\ket{11110000}+\ket{10110001}+\ket{01110010}+\ket{00110011}\\
       &+\ket{01110100}+\ket{00110101}+\ket{11110110}+\ket{10110111}\\
       &+\ket{11101000}+\ket{10101001}+\ket{01101010}+\ket{00101011}\\
       &+\ket{01101100}+\ket{00101101}+\ket{11101110}+\ket{10101111}\\
       &+\ket{10111000}+\ket{11111001}+\ket{00111010}+\ket{01111011}\\
       &+\ket{00111100}+\ket{01111101}+\ket{10111110}+\ket{11111111}).
       \label{eq33}
    \end{split}
\end{equation}
\endgroup

To obtain the remotely prepared state, we perform measurements over all position states and selected coin spaces. If the measurement outcomes for both Alice’s and Bob’s position bases are  $\ket{\alpha_{0}}=\ket{00}$, then Eq.~(\ref{eq33}) becomes,

 \begin{equation}
 \begin{split}
     \ket{\psi_{7}}=&\frac{1}{8}(\ket{000000}+\ket{000110}+\ket{010000}+\ket{010011}\\
     &+\ket{001001}+\ket{001111}+\ket{011000}+\ket{011110}\\
     &+\ket{100010}+\ket{100100}+\ket{110011}+\ket{110101}\\
     &+\ket{101011}+\ket{101101}+\ket{111010}+\ket{111100}).
     \end{split}
 \end{equation}

 Then, Alice and Bob ($A_2$ and $B_2$) perform measurements in the  $\ket{\beta_0}\ket{\gamma_0}$ bases, as defined in Eqs.~(\ref{eq46}) and (\ref{eq47}) yielding,

 \begin{equation}
 \begin{split}
     \ket{\psi_{8}}=&a_{0}b_{0}\ket{0000}+a_{0}b_{0}\ket{0110}+a_{0}b_{0}\ket{1000}+a_{0}b_{0}\ket{1011}\\
     &+a_{0}b_{1}\ket{0001}+a_{0}b_{1}\ket{0111}+a_{0}b_{1}\ket{1000}+a_{0}b_{1}\ket{1110}\\
     &+a_{1}b_{0}\ket{0010}+a_{1}b_{0}\ket{0100}+a_{1}b_{0}\ket{1011}+a_{1}b_{0}\ket{1101}\\
     &+a_{1}b_{1}\ket{0011}+a_{1}b_{1}\ket{0101}+a_{1}b_{1}\ket{1010}+a_{1}b_{1}\ket{1100}.
     \end{split}
 \end{equation}

Finally, Charlie performs measurements on particles $C_1$ and $C_2$ in the $\ket{00}$ basis, yielding,

\begin{equation}
 \begin{split}
     \ket{\psi_{9}}=(b_{0}\ket{0}+b_{1}\ket{1})_{A_{3}}(a_{0}\ket{0}+a_{1}\ket{1})_{B_{3}}
     \end{split}
 \end{equation}

with a probability of $\frac{1}{64}$. That is, Alice is able to create a known state for Bob, and Bob is able to create a known state for Alice simultaneously. Alice and Bob can perform measurements in all possible position bases $(~ \ket{\alpha_1}=\ket{01},~ \ket{\alpha_2}=\ket{10},~ \ket{\alpha_3}=\ket{11})$
and coin bases, followed by Charlie’s measurement in the coin bases, which completes the protocol as shown Appendix B, Table \ref{table3}. 

\subsection{Controlled Bidirectional Remote State Preparation Using Quantum Walks on 4-Cycle}

In this controlled BRSP scheme, the position spaces of Alice and Bob are two independent 4-cycles, as shown in FIG. \ref{fig:fig2} of the previous section. In addition to the uncontrolled case, two more coin spaces, $C_1$ and $C_2$, are introduced by the controller. The conditional shift operator in this case is the same as that in Eqn.~(\ref{eq25}) of Sect.~I.

Initially, the position spaces of $A_1$ and $B_1$ are initialized to $\ket{0}$, and the coin spaces of $A_2$,$A_3$,$B_2$,$B_3$, $C_1$and$C_2$ are initialized to $\ket{0}$. Hence, the total initial state, in a convenient order, is,

\begin{equation}
    \ket{\psi_{0}}_{A_{1}B_{1}A_{2}A_{3}B_{2}B_{3}C_{1}C_{2}}=\ket{00}\otimes\ket{0}\otimes\ket{0}\otimes\ket{0}\otimes\ket{0}\otimes\ket{0}\otimes\ket{0}.
    \label{eq39}
 \end{equation}

As in the previous section, we perform all six steps of the quantum walk. Now, we analyze the protocol in detail by first applying the initial step of the quantum walk, $W_1$, to Eq.~(\ref{eq39}),

  \begin{equation}
    \ket{\psi_{1}}=\frac{1}{\sqrt{2}}(\ket{10000000}+\ket{30100000}).
    \label{Eq2}
 \end{equation}

Then, applying the second step, $W_2$, we obtain the state,
 
 \begin{equation}
    \ket{\psi_{2}}=\frac{1}{2}(\ket{11000000}+\ket{13001000}+\ket{31100000}+\ket{33101000}).
    \label{Eq2}
 \end{equation}

Following the third step of the quantum walk, $W_3$, the above state is transformed to,
\begin{equation}
\begin{split}
\ket{\psi_{3}}=&\frac{1}{2}(\ket{12000000}+\ket{10010000}+\ket{10001000}+\ket{12011000} \\&+\ket{32100000}+\ket{30110000}+\ket{30101000}+\ket{32101000}).  
\end{split}
\end{equation}

Next, upon applying the fourth step, $W_4$, we obtain the state,
\begingroup
\allowdisplaybreaks
\begin{equation}
\begin{split}
\ket{\psi_{4}}=&\frac{1}{4}(\ket{22000000}+\ket{02000100}+\ket{20010000}+\ket{00010100}\\
&+\ket{20001000}+\ket{00001100}+\ket{22011000}+\ket{02011100}\\ &+\ket{02100000}+\ket{22100100}+\ket{00110000}+\ket{20110100}\\
&+\ket{00101000}+\ket{20101100}+\ket{02111000}+\ket{22111100}).   
\end{split}
\end{equation}
\endgroup

After applying the fifth step, $W_5$, to the above state, we obtain,

\begin{equation}
    \begin{split}
       \ket{\psi_{5}}=&\frac{1}{4\sqrt{2}}(\ket{32000000}+\ket{12000010}
    +\ket{12000100}+\ket{32000110}\\
       &+\ket{30010000}+\ket{10010010}+\ket{10010100}+\ket{30010110}+\\
       &+\ket{30001000}+\ket{10001010}+\ket{10001100}+\ket{30001110}+\\
       &+\ket{32011000}+\ket{12011010}+\ket{12011100}+\ket{32011110}\\
       &+\ket{12100000}+\ket{32100010}+\ket{32100100}+\ket{12100110}\\
       &+\ket{10110000}+\ket{30110010}+\ket{30110100}+\ket{10110110}+\\
       &+\ket{10101000}+\ket{30101010}+\ket{30101100}+\ket{10101110}\\
       &+\ket{12111000}+\ket{32111010}+\ket{32111100}+\ket{12111110}).
    \end{split}
\end{equation}

Finally, upon applying $W_6$, the final state is obtained,

\begin{equation}
    \begin{split}
       \ket{\psi_{6}}=&\frac{1}{8}(\ket{33000000}+\ket{31000001}+\ket{13000010}+\ket{11000011}\\
       &+\ket{13000100}+\ket{11000101}+\ket{33000110}+\ket{31000111}\\
       &+\ket{31010000}+\ket{33010001}+\ket{11010010}+\ket{13010011}\\
       &+\ket{11010100}+\ket{13010101}+\ket{31010110}+\ket{33010111}\\
       &+\ket{31001000}+\ket{33001001}+\ket{11001010}+\ket{13001011}\\
       &+\ket{11001100}+\ket{13001101}+\ket{31001110}+\ket{33001111}\\
       &+\ket{33011000}+\ket{31011001}+\ket{13011010}+\ket{11011011}\\
       &+\ket{13011100}+\ket{11011101}+\ket{33011110}+\ket{31011111}\\
       &+\ket{13100000}+\ket{11100001}+\ket{33100010}+\ket{31100011}\\
       &+\ket{33100100}+\ket{31100101}+\ket{13100110}+\ket{11100111}\\
       &+\ket{11110000}+\ket{13110001}+\ket{31110010}+\ket{33110011}\\
       &+\ket{31110100}+\ket{33110101}+\ket{11110110}+\ket{13110111}\\
       &+\ket{11101000}+\ket{13101001}+\ket{31101010}+\ket{33101011}\\
       &+\ket{31101100}+\ket{33101101}+\ket{11101110}+\ket{13101111}\\
       &+\ket{13111000}+\ket{11111001}+\ket{33111010}+\ket{31111011}\\
       &+\ket{33111100}+\ket{31111101}+\ket{13111110}+\ket{11111111}).
       \label{eq38}
    \end{split}
\end{equation}

To obtain the remotely prepared state, we perform measurements over all position states and select the relevant components of the coin states. If the measurement outcome corresponds to both Alice’s and Bob’s position basis being  $\ket{\tilde{\alpha_{0}}}=\ket{33}$, then Eq.~(\ref{eq38}) becomes,

 \begin{equation}
 \begin{split}
     \ket{\psi_{7}}=&\frac{1}{8}(\ket{000000}+\ket{000110}+\ket{010000}+\ket{010011}\\
     &+\ket{001001}+\ket{001111}+\ket{011000}+\ket{011110}\\
     &+\ket{100010}+\ket{100100}+\ket{110011}+\ket{110101}\\
     &+\ket{101011}+\ket{101101}+\ket{111010}+\ket{111100}).
     \end{split}
 \end{equation}

 Then, particles $A_2$ and $B_2$  are measured in the $\ket{\beta_0}\ket{\gamma_0}$ bases defined in Eqs.~(\ref{eq46}) and (\ref{eq47}), yielding,

 \begin{equation}
 \begin{split}
     \ket{\psi_{8}}=&a_{0}b_{0}\ket{0000}+a_{0}b_{0}\ket{0110}+a_{0}b_{0}\ket{1000}+a_{0}b_{0}\ket{1011}\\
     &+a_{0}b_{1}\ket{0001}+a_{0}b_{1}\ket{0111}+a_{0}b_{1}\ket{1000}+a_{0}b_{1}\ket{1110}\\
     &+a_{1}b_{0}\ket{0010}+a_{1}b_{0}\ket{0100}+a_{1}b_{0}\ket{1011}+a_{1}b_{0}\ket{1101}\\
     &+a_{1}b_{1}\ket{0011}+a_{1}b_{1}\ket{0101}+a_{1}b_{1}\ket{1010}+a_{1}b_{1}\ket{1100}.
     \end{split}
 \end{equation}

Finally, Charlie performs measurements on coins $C_1$ and $C_2$ in the $\ket{00}$ basis, yielding,

\begin{equation}
 \begin{split}
     \ket{\psi_{9}}=(b_{0}\ket{0}+b_{1}\ket{1})_{A_{3}}(a_{0}\ket{0}+a_{1}\ket{1})_{B_{3}}
     \end{split}
 \end{equation}

with a probability of $\frac{1}{64}$. That is, Alice is able to create a known state for Bob, and Bob is able to create a known state for Alice simultaneously.  Instead of measuring the above-mentioned position basis, $\ket{\tilde{\alpha_{0}}}=\ket{33}$, Alice and Bob can measure the other position bases $\ket{\tilde{\alpha_{1}}} =\ket{31}$, $ \ket{\tilde{\alpha_{2}}}=\ket{13}$ and $ \ket{\tilde{\alpha_{3}}}=\ket{11}$ followed by measurements in all possible coin bases. The obtained results are equivalent to the controlled BRSP in the two-vertex case.  The corresponding table for this case is analogous to Table. \ref{table3} in Appendix B, with the position basis $\ket{\alpha_{i}}$  in the table replaced by $\ket{\tilde{\alpha_{i}}}$.   
\section{Conclusion \label{sec4}}
We investigated bidirectional remote state preparation using quantum walks on two independent one-dimensional lattices, as well as on two independent cyclic graphs with two and four vertices. In this framework, nearest-neighbor transitions controlled by coin operations were used to achieve remote preparation of quantum states. The protocol was implemented in two scenarios: one without a controller and one with the involvement of a controller.

In this method, the entanglement required between the communicating parties is generated dynamically during the quantum walk evolution, eliminating the need for any pre-shared entangled resource. We analyzed the bidirectional remote state preparation process for both the uncontrolled and controlled cases on the considered graph structures. Finally, we showed that the quantum walk–based schemes on the two-vertex and four-cycle graphs display consistent behavior in both uncontrolled and controlled configurations.
\appendix
\section{Complete table for Controlled Bidirectional Remote State Preparation protocol}
The controlled bidirectional remote state preparation protocol on a line is summarized in the following table.
\begin{center}
\captionof{table}{Outcome of measurement performed by Alice and Bob in the first coin state and the unitary operation on their second coin state and measurement performed by  Charlie in his two coin state for creating the known single qubit state.}
\begin{supertabular}{p{3cm} p{4cm} p{3cm} p{5cm}}
\hline
~~~Position basis~~~ &  ~~~~ $A_{2},B_{2}$ Coin basis~~~~ &$C_{1},C_{2}$ Coin basis  ~~~& ~~~~~~ Unitary operation ~~~~~~ \\
\hline
\multirow{8}{*}{$\ket{\alpha_0}_{A_{1}B_{1}}$} & \multirow{4}{*}{$\ket{\beta_{0}}_{A_{2}}\ket{\gamma_{0}}_{B_{2}}$} & $\ket{00}$ & I \\
&& $\ket{01} $ &  $\sigma_{A_{3}}^{x}$\\
&& $\ket{10} $ &  $\sigma_{B_{3}}^{x}$\\
&& $\ket{11} $ &  $\sigma_{A_{3}}^{x} \sigma_{B_{3}}^{x} $ \\
\cline{2-4}
& $\ket{\beta_{0}}_{A_{2}}\ket{\gamma_{1}}_{B_{2}}$ & $\ket{00}$ & $\sigma_{A_{3}}^{x} \sigma_{A_{3}}^{z}$ \\
&& $\ket{01} $ &  $\sigma_{A_{3}}^{x}\sigma_{A_{3}}^{z} \sigma_{A_{3}}^{x}$\\
&& $\ket{10} $ &  $\sigma_{A_{3}}^{x}\sigma_{A_{3}}^{z} \sigma_{B_{3}}^{x}$\\
&& $\ket{11} $ &  $\sigma_{A_{3}}^{x} \sigma_{A_{3}}^{z} \sigma_{A_{3}}^{x} \sigma_{B_{3}}^{x}$ \\
\cline{2-4}
& \multirow{4}{*}{$\ket{\beta_{1}}_{A_{2}}\ket{\gamma_{0}}_{B_{2}}$} & $\ket{00}$ & $\sigma_{B_{3}}^{x} \sigma_{B_{3}}^{z}$ \\
&& $\ket{01} $ &  $\sigma_{A_{3}}^{x} \sigma_{B_{3}}^{x} \sigma_{B_{3}}^{z} $\\
&& $\ket{10} $ &  $ \sigma_{B_{3}}^{x} \sigma_{B_{3}}^{z} \sigma_{B_{3}}^{x} $\\
&& $\ket{11} $ &  $\sigma_{A_{3}}^{x} \sigma_{B_{3}}^{x} \sigma_{B_{3}}^{z} \sigma_{B_{3}}^{x}$\\
\cline{2-4}
& \multirow{4}{*}{$\ket{\beta_{1}}_{A_{2}}\ket{\gamma_{1}}_{B_{2}}$} & $\ket{00}$ & $\sigma_{A_{3}}^{x} \sigma_{A_{3}}^{z} \sigma_{B_{3}}^{x} \sigma_{B_{3}}^{z}$ \\
&& $\ket{01}$ & $\sigma_{A_{3}}^{x} \sigma_{A_{3}}^{z} \sigma_{A_{3}}^{x} \sigma_{B_{3}}^{x} \sigma_{B_{3}}^{z}$ \\
&& $\ket{10}$ & $\sigma_{A_{3}}^{x} \sigma_{A_{3}}^{z}  \sigma_{B_{3}}^{x} \sigma_{B_{3}}^{z} \sigma_{B_{3}}^{x}$ \\
&& $\ket{11}$ & $\sigma_{A_{3}}^{x} \sigma_{A_{3}}^{z} \sigma_{A_{3}}^{x} \sigma_{B_{3}}^{x} \sigma_{B_{3}}^{z} \sigma_{B_{3}}^{x}$ \\
[1.ex]
\hline
\multirow{16}{*}{$\ket{\alpha_1}_{A_{1}B_{1}}$} & \multirow{4}{*}{$\ket{\beta_{0}}_{A_{2}}\ket{\gamma_{0}}_{B_{2}}$} & $\ket{00}$ & $\sigma_{A_{3}}^{z}$ \\
&& $\ket{01} $ &  $\sigma_{A_{3}}^{z}\sigma_{A_{3}}^{x}\sigma_{A_{3}}^{z}$\\
&& $\ket{10} $ &  $\sigma_{A_{3}}^{z}\sigma_{B_{3}}^{x}$\\
&& $\ket{11} $ &  $\sigma_{A_{3}}^{z}\sigma_{A_{3}}^{x}\sigma_{A_{3}}^{z} \sigma_{B_{3}}^{x} $ \\
\cline{2-4}
& \multirow{4}{*}{$\ket{\beta_{0}}_{A_{2}}\ket{\gamma_{1}}_{B_{2}}$} & $\ket{00}$ & $\sigma_{A_{3}}^{x} $ \\
&& $\ket{01} $ &  $\sigma_{A_{3}}^{z} $\\
&& $\ket{10} $ &  $\sigma_{A_{3}}^{x} \sigma_{B_{3}}^{x}$\\
&& $\ket{11} $ &  $\sigma_{A_{3}}^{z}  \sigma_{B_{3}}^{x}$ \\
\cline{2-4}
& \multirow{4}{*}{$\ket{\beta_{1}}_{A_{2}}\ket{\gamma_{0}}_{B_{2}}$} & $\ket{00}$ & $\sigma_{A_{3}}^{x}\sigma_{A_{3}}^{z}\sigma_{A_{3}}^{x}\sigma_{B_{3}}^{z} \sigma_{B_{3}}^{x}$ \\
&& $\ket{01} $ &  $\sigma_{A_{3}}^{x} \sigma_{B_{3}}^{z} \sigma_{B_{3}}^{x} $\\
&& $\ket{10} $ &  $ \sigma_{A_{3}}^{x}\sigma_{A_{3}}^{z}\sigma_{A_{3}}^{x}\sigma_{B_{3}}^{z}$\\
&& $\ket{11} $ &  $\sigma_{A_{3}}^{x}  \sigma_{B_{3}}^{z} $\\
\cline{2-4}
& \multirow{4}{*}{$\ket{\beta_{1}}_{A_{2}}\ket{\gamma_{1}}_{B_{2}}$} & $\ket{00}$ & $\sigma_{A_{3}}^{x}  \sigma_{B_{3}}^{x} \sigma_{B_{3}}^{z}$ \\
&& $\ket{01}$ & $ \sigma_{A_{3}}^{z} \sigma_{B_{3}}^{x} \sigma_{B_{3}}^{z}$ \\
&& $\ket{10}$ & $\sigma_{A_{3}}^{x}   \sigma_{B_{3}}^{x} \sigma_{B_{3}}^{z} \sigma_{B_{3}}^{x}$ \\
&& $\ket{11}$ & $\sigma_{A_{3}}^{z}  \sigma_{B_{3}}^{x} \sigma_{B_{3}}^{z} \sigma_{B_{3}}^{x}$ \\
[1.ex]
\hline
 \multirow{16}{*}{$\ket{\alpha_2}_{A_{1}B_{1}}$} &  \multirow{4}{*}{$\ket{\beta_{0}}_{A_{2}}\ket{\gamma_{0}}_{B_{2}}$} & $\ket{00}$ & $\sigma_{B_{3}}^{z}$ \\
&& $\ket{01} $ &  $\sigma_{A_{3}}^{x}\sigma_{B_{3}}^{z}$\\
&& $\ket{10} $ &  $\sigma_{B_{3}}^{z}\sigma_{B_{3}}^{x}\sigma_{B_{3}}^{z}$\\
&& $\ket{11} $ &  $\sigma_{A_{3}}^{x} \sigma_{B_{3}}^{z}\sigma_{B_{3}}^{x}\sigma_{B_{3}}^{z} $ \\
\cline{2-4}
&  \multirow{4}{*}{$\ket{\beta_{0}}_{A_{2}}\ket{\gamma_{1}}_{B_{2}}$} & $\ket{00}$ & $\sigma_{A_{3}}^{z}\sigma_{A_{3}}^{x}\sigma_{B_{3}}^{x}\sigma_{B_{3}}^{z} $ \\
&& $\ket{01} $ &  $\sigma_{A_{3}}^{z}\sigma_{B_{3}}^{x}\sigma_{B_{3}}^{z} $\\
&& $\ket{10} $ &  $\sigma_{A_{3}}^{z}\sigma_{A_{3}}^{x} \sigma_{B_{3}}^{x}$\\
&& $\ket{11} $ &  $\sigma_{A_{3}}^{z}  \sigma_{B_{3}}^{x}$ \\
\cline{2-4}
&  $\ket{\beta_{1}}_{A_{2}}\ket{\gamma_{0}}_{B_{2}}$ & $\ket{00}$ & $ \sigma_{B_{3}}^{x}$ \\
&& $\ket{01} $ &  $\sigma_{A_{3}}^{x}  \sigma_{B_{3}}^{x} $\\
&& $\ket{10} $ &  $ \sigma_{B_{3}}^{z}$\\
&& $\ket{11} $ &  $\sigma_{A_{3}}^{x}  \sigma_{B_{3}}^{z} $\\
\cline{2-4}
&  \multirow{4}{*}{$\ket{\beta_{1}}_{A_{2}}\ket{\gamma_{1}}_{B_{2}}$} & $\ket{00}$ & $\sigma_{A_{3}}^{x} \sigma_{A_{3}}^{z} \sigma_{B_{3}}^{x} $ \\
&& $\ket{01}$ & $\sigma_{A_{3}}^{x} \sigma_{A_{3}}^{z}\sigma_{A_{3}}^{x} \sigma_{B_{3}}^{x} $ \\
&& $\ket{10}$ & $\sigma_{A_{3}}^{x} \sigma_{A_{3}}^{z}   \sigma_{B_{3}}^{z} $ \\
&& $\ket{11}$ & $\sigma_{A_{3}}^{x} \sigma_{A_{3}}^{z}\sigma_{A_{3}}^{x} \sigma_{B_{3}}^{z}$ \\
[1.ex]
\hline

 \multirow{16}{*}{$\ket{\alpha_3}_{A_{1}B_{1}}$} &  \multirow{4}{*}{$\ket{\beta_{0}}_{A_{2}}\ket{\gamma_{0}}_{B_{2}}$} & $\ket{00}$ & $\sigma_{A_{3}}^{x}\sigma_{A_{3}}^{z}\sigma_{A_{3}}^{x}\sigma_{B_{3}}^{x}\sigma_{B_{3}}^{z}\sigma_{B_{3}}^{x}$ \\
&& $\ket{01} $ &  $\sigma_{A_{3}}^{x}\sigma_{B_{3}}^{x}\sigma_{B_{3}}^{z}\sigma_{B_{3}}^{x}$\\
&& $\ket{10} $ &  $\sigma_{A_{3}}^{x}\sigma_{A_{3}}^{z}\sigma_{A_{3}}^{x}\sigma_{B_{3}}^{x}$\\
&& $\ket{11} $ &  $\sigma_{A_{3}}^{x} \sigma_{B_{3}}^{x} $ \\
\cline{2-4}
&  \multirow{4}{*}{$\ket{\beta_{0}}_{A_{2}}\ket{\gamma_{1}}_{B_{2}}$} & $\ket{00}$ & $\sigma_{A_{3}}^{x}\sigma_{B_{3}}^{z} $ \\
&& $\ket{01} $ &  $\sigma_{A_{3}}^{z}\sigma_{B_{3}}^{z} $\\
&& $\ket{10} $ &  $\sigma_{A_{3}}^{x} \sigma_{B_{3}}^{z}\sigma_{B_{3}}^{x}\sigma_{B_{3}}^{z}$\\
&& $\ket{11} $ &  $\sigma_{A_{3}}^{z} \sigma_{B_{3}}^{z} \sigma_{B_{3}}^{x}\sigma_{B_{3}}^{z}$ \\
\cline{2-4}
&  \multirow{4}{*}{$\ket{\beta_{1}}_{A_{2}}\ket{\gamma_{0}}_{B_{2}}$} & $\ket{00}$ & $ \sigma_{A_{3}}^{z}\sigma_{B_{3}}^{x}$ \\
&& $\ket{01} $ &  $\sigma_{A_{3}}^{z}\sigma_{A_{3}}^{x}\sigma_{A_{3}}^{z}  \sigma_{B_{3}}^{x} $\\
&& $\ket{10} $ &  $ \sigma_{A_{3}}^{z}\sigma_{B_{3}}^{z}$\\
&& $\ket{11} $ &  $\sigma_{A_{3}}^{z}\sigma_{A_{3}}^{x}\sigma_{A_{3}}^{z}  \sigma_{B_{3}}^{z} $\\
\cline{2-4}
&  \multirow{4}{*}{$\ket{\beta_{1}}_{A_{2}}\ket{\gamma_{1}}_{B_{2}}$} & $\ket{00}$ & $\sigma_{A_{3}}^{x}  \sigma_{B_{3}}^{x} $ \\
&& $\ket{01}$ & $ \sigma_{A_{3}}^{z}\sigma_{B_{3}}^{x} $ \\
&& $\ket{10}$ & $\sigma_{A_{3}}^{x}   \sigma_{B_{3}}^{z} $ \\
&& $\ket{11}$ & $ \sigma_{A_{3}}^{z} \sigma_{B_{3}}^{z}$ \\
\hline

\multirow{16}{*}{$\ket{\alpha_4}_{A_{1}B_{1}}$} & \multirow{4}{*}{$\ket{\beta_{0}}_{A_{2}}\ket{\gamma_{0}}_{B_{2}}$} & $\ket{00}$ & $\sigma_{A_{3}}^{x} $\\
&& $\ket{01}$ & I \\
&& $\ket{10} $ &  $\sigma_{A_{3}}^{x} \sigma_{B_{3}}^{x} $\\
&& $\ket{11} $ &  $ \sigma_{B_{3}}^{x} $\\
\cline{2-4}
& \multirow{4}{*}{$\ket{\beta_{0}}_{A_{2}}\ket{\gamma_{1}}_{B_{2}}$} & $\ket{00}$ & $\sigma_{A_{3}}^{x} \sigma_{A_{3}}^{z} \sigma_{A_{3}}^{x}$ \\
&& $\ket{01}$ & $\sigma_{A_{3}}^{x} \sigma_{A_{3}}^{z} $ \\
&& $\ket{10}$ & $\sigma_{A_{3}}^{x} \sigma_{A_{3}}^{z} \sigma_{A_{3}}^{x} \sigma_{B_{3}}^{x}$ \\
&& $\ket{11}$ & $\sigma_{A_{3}}^{x} \sigma_{A_{3}}^{z}  \sigma_{B_{3}}^{x}$\\
\cline{2-4}
& \multirow{4}{*}{$\ket{\beta_{1}}_{A_{2}}\ket{\gamma_{0}}_{B_{2}}$} & $\ket{00}$ & $\sigma_{A_{3}}^{x}  \sigma_{B_{3}}^{x} \sigma_{B_{3}}^{z}$ \\
&& $\ket{01}$ & $\sigma_{B_{3}}^{x} \sigma_{B_{3}}^{z} $ \\
&& $\ket{10}$ & $ \sigma_{A_{3}}^{x}\sigma_{B_{3}}^{x} \sigma_{B_{3}}^{z} \sigma_{B_{3}}^{x} $ \\
&& $\ket{11}$ & $ \sigma_{B_{3}}^{x} \sigma_{B_{3}}^{z} \sigma_{B_{3}}^{x} $ \\
\cline{2-4}
&$\ket{\beta_{1}}_{A_{2}}\ket{\gamma_{1}}_{B_{2}}$ & $\ket{00}$ & $\sigma_{A_{3}}^{x} \sigma_{A_{3}}^{z}\sigma_{A_{3}}^{x} \sigma_{B_{3}}^{x} \sigma_{B_{3}}^{z}$ \\
&& $\ket{01}$ & $\sigma_{A_{3}}^{x} \sigma_{A_{3}}^{z} \sigma_{B_{3}}^{x} \sigma_{B_{3}}^{z}$ \\
&& $\ket{10}$ & $\sigma_{A_{3}}^{x} \sigma_{A_{3}}^{z}\sigma_{A_{3}}^{x} \sigma_{B_{3}}^{x} \sigma_{B_{3}}^{z} \sigma_{B_{3}}^{x}$ \\
&& $\ket{11}$ & $\sigma_{A_{3}}^{x} \sigma_{A_{3}}^{z} \sigma_{B_{3}}^{x} \sigma_{B_{3}}^{z} \sigma_{B_{3}}^{x}$ \\
[1.ex]
\hline

\multirow{16}{*}{$\ket{\alpha_5}_{A_{1}B_{1}}$} & \multirow{4}{*}{$\ket{\beta_{0}}_{A_{2}}\ket{\gamma_{0}}_{B_{2}}$} & $\ket{00}$ & $\sigma_{A_{3}}^{x} $\\
&& $\ket{01}$ & $\sigma_{A_{3}}^{z}$ \\
&& $\ket{10} $ &  $\sigma_{A_{3}}^{x} \sigma_{B_{3}}^{x} $\\
&& $\ket{11} $ &  $ \sigma_{A_{3}}^{z}\sigma_{B_{3}}^{x} $\\
\cline{2-4}
& \multirow{4}{*}{$\ket{\beta_{0}}_{A_{2}}\ket{\gamma_{1}}_{B_{2}}$} & $\ket{00}$ & $\sigma_{A_{3}}^{x} \sigma_{A_{3}}^{z} \sigma_{A_{3}}^{x}$ \\
&& $\ket{01}$ & $\sigma_{A_{3}}^{x}  $ \\
&& $\ket{10}$ & $\sigma_{A_{3}}^{x} \sigma_{A_{3}}^{z} \sigma_{A_{3}}^{x} \sigma_{B_{3}}^{x}$ \\
&& $\ket{11}$ & $\sigma_{A_{3}}^{x}  \sigma_{B_{3}}^{x}$\\
\cline{2-4}
& \multirow{4}{*}{$\ket{\beta_{1}}_{A_{2}}\ket{\gamma_{0}}_{B_{2}}$} & $\ket{00}$ & $\sigma_{A_{3}}^{x}  \sigma_{B_{3}}^{x} \sigma_{B_{3}}^{z}$ \\
&& $\ket{01}$ & $\sigma_{A_{3}}^{z}\sigma_{B_{3}}^{x} \sigma_{B_{3}}^{z} $ \\
&& $\ket{10}$ & $ \sigma_{A_{3}}^{x}\sigma_{B_{3}}^{x} \sigma_{B_{3}}^{z} \sigma_{B_{3}}^{x} $ \\
&& $\ket{11}$ & $ \sigma_{A_{3}}^{z}\sigma_{B_{3}}^{x} \sigma_{B_{3}}^{z} \sigma_{B_{3}}^{x} $ \\
\cline{2-4}
& \multirow{4}{*}{$\ket{\beta_{1}}_{A_{2}}\ket{\gamma_{1}}_{B_{2}}$} & $\ket{00}$ & $\sigma_{A_{3}}^{x} \sigma_{A_{3}}^{z}\sigma_{A_{3}}^{x} \sigma_{B_{3}}^{x} \sigma_{B_{3}}^{z}$ \\
&& $\ket{01}$ & $\sigma_{A_{3}}^{x} \sigma_{B_{3}}^{x} \sigma_{B_{3}}^{z}$ \\
&& $\ket{10}$ & $\sigma_{A_{3}}^{x} \sigma_{A_{3}}^{z}\sigma_{A_{3}}^{x} \sigma_{B_{3}}^{x} \sigma_{B_{3}}^{z} \sigma_{B_{3}}^{x}$ \\
&& $\ket{11}$ & $\sigma_{A_{3}}^{x}  \sigma_{B_{3}}^{x} \sigma_{B_{3}}^{z} \sigma_{B_{3}}^{x}$ \\
[1.ex]
\hline

\multirow{16}{*}{$\ket{\alpha_6}_{A_{1}B_{1}}$} & \multirow{4}{*}{$\ket{\beta_{0}}_{A_{2}}\ket{\gamma_{0}}_{B_{2}}$} & $\ket{00}$ & $\sigma_{A_{3}}^{x}\sigma_{B_{3}}^{z} $\\
&& $\ket{01}$ & $\sigma_{B_{3}}^{z}$ \\
&& $\ket{10} $ &  $\sigma_{A_{3}}^{x} \sigma_{B_{3}}^{z}\sigma_{B_{3}}^{x}\sigma_{B_{3}}^{z} $\\
&& $\ket{11} $ &  $ \sigma_{B_{3}}^{z}\sigma_{B_{3}}^{x} \sigma_{B_{3}}^{z}$\\
\cline{2-4}
& \multirow{4}{*}{$\ket{\beta_{0}}_{A_{2}}\ket{\gamma_{1}}_{B_{2}}$} & $\ket{00}$ & $ \sigma_{A_{3}}^{z}\sigma_{B_{3}}^{x}\sigma_{B_{3}}^{z} \sigma_{B_{3}}^{x}$ \\
&& $\ket{01}$ & $ \sigma_{A_{3}}^{z}\sigma_{A_{3}}^{x}\sigma_{B_{3}}^{x}\sigma_{B_{3}}^{z} \sigma_{B_{3}}^{x} $ \\
&& $\ket{10}$ & $ \sigma_{A_{3}}^{z} \sigma_{B_{3}}^{x}$ \\
&& $\ket{11}$ & $ \sigma_{A_{3}}^{z}\sigma_{A_{3}}^{x}  \sigma_{B_{3}}^{x}$\\
\cline{2-4}
& \multirow{4}{*}{$\ket{\beta_{1}}_{A_{2}}\ket{\gamma_{0}}_{B_{2}}$} & $\ket{00}$ & $\sigma_{A_{3}}^{x}  \sigma_{B_{3}}^{x} $ \\
&& $\ket{01}$ & $\sigma_{B_{3}}^{x}  $ \\
&& $\ket{10}$ & $ \sigma_{A_{3}}^{x} \sigma_{B_{3}}^{z}  $ \\
&& $\ket{11}$ & $  \sigma_{B_{3}}^{z} $ \\
\cline{2-4}
& \multirow{4}{*}{$\ket{\beta_{1}}_{A_{2}}\ket{\gamma_{1}}_{B_{2}}$} & $\ket{00}$ & $\sigma_{A_{3}}^{x} \sigma_{A_{3}}^{z}\sigma_{A_{3}}^{x} \sigma_{B_{3}}^{x} $ \\
&& $\ket{01}$ & $\sigma_{A_{3}}^{x} \sigma_{A_{3}}^{z} \sigma_{B_{3}}^{x} $ \\
&& $\ket{10}$ & $\sigma_{A_{3}}^{x} \sigma_{A_{3}}^{z}\sigma_{A_{3}}^{x}  \sigma_{B_{3}}^{z} $ \\
&& $\ket{11}$ & $\sigma_{A_{3}}^{x} \sigma_{A_{3}}^{z}  \sigma_{B_{3}}^{z} $ \\
\hline

$\ket{\alpha_7}_{A_{1}B_{1}}$ & $\ket{\beta_{0}}_{A_{2}}\ket{\gamma_{0}}_{B_{2}}$ & $\ket{00}$ & $\sigma_{A_{3}}^{x}\sigma_{B_{3}}^{z} $\\
&& $\ket{01}$ & $\sigma_{A_{3}}^{z}\sigma_{B_{3}}^{z}$ \\
&& $\ket{10} $ &  $\sigma_{A_{3}}^{x} \sigma_{B_{3}}^{z}\sigma_{B_{3}}^{x}\sigma_{B_{3}}^{z} $\\
&& $\ket{11} $ &  $ \sigma_{A_{3}}^{z}\sigma_{B_{3}}^{z}\sigma_{B_{3}}^{x} \sigma_{B_{3}}^{z}$\\
\cline{2-4}
& \multirow{4}{*}{$\ket{\beta_{0}}_{A_{2}}\ket{\gamma_{1}}_{B_{2}}$} & $\ket{00}$ & $ \sigma_{A_{3}}^{z}\sigma_{A_{3}}^{x}\sigma_{B_{3}}^{z} $ \\
&& $\ket{01}$ & $ \sigma_{A_{3}}^{x}\sigma_{B_{3}}^{z}  $ \\
&& $\ket{10}$ & $ \sigma_{A_{3}}^{z}\sigma_{A_{3}}^{x} \sigma_{B_{3}}^{z}\sigma_{B_{3}}^{x}\sigma_{B_{3}}^{z}$ \\
&& $\ket{11}$ & $ \sigma_{A_{3}}^{x}  \sigma_{B_{3}}^{z}\sigma_{B_{3}}^{x}\sigma_{B_{3}}^{z}$\\
\cline{2-4}
& \multirow{4}{*}{$\ket{\beta_{1}}_{A_{2}}\ket{\gamma_{0}}_{B_{2}}$} & $\ket{00}$ & $\sigma_{A_{3}}^{x}  \sigma_{B_{3}}^{x} $ \\
&& $\ket{01}$ & $\sigma_{A_{3}}^{z}\sigma_{B_{3}}^{x}  $ \\
&& $\ket{10}$ & $ \sigma_{A_{3}}^{x} \sigma_{B_{3}}^{z}  $ \\
&& $\ket{11}$ & $  \sigma_{A_{3}}^{z}\sigma_{B_{3}}^{z} $ \\
\cline{2-4}
& \multirow{4}{*}{$\ket{\beta_{1}}_{A_{2}}\ket{\gamma_{1}}_{B_{2}}$} & $\ket{00}$ & $\sigma_{A_{3}}^{x} \sigma_{A_{3}}^{z}\sigma_{A_{3}}^{x} \sigma_{B_{3}}^{x} $ \\
&& $\ket{01}$ & $\sigma_{A_{3}}^{x} \sigma_{B_{3}}^{z} $ \\
&& $\ket{10}$ & $\sigma_{A_{3}}^{x} \sigma_{A_{3}}^{z}\sigma_{A_{3}}^{x}  \sigma_{B_{3}}^{z} $ \\
&& $\ket{11}$ & $\sigma_{A_{3}}^{x}   \sigma_{B_{3}}^{z} $ \\
[1.ex]
\hline

\multirow{16}{*}{$\ket{\alpha_8}_{A_{1}B_{1}}$} &  \multirow{4}{*}{$\ket{\beta_{0}}_{A_{2}}\ket{\gamma_{0}}_{B_{2}}$} & $\ket{00}$ & $\sigma_{B_{3}}^{x} $\\
&& $\ket{01}$ & $\sigma_{A_{3}}^{x} \sigma_{B_{3}}^{x} $ \\
&& $\ket{10} $ &  I\\
&& $\ket{11} $ &  $ \sigma_{A_{3}}^{x} $\\
\cline{2-4}
&  \multirow{4}{*}{$\ket{\beta_{0}}_{A_{2}}\ket{\gamma_{1}}_{B_{2}}$} & $\ket{00}$ & $\sigma_{A_{3}}^{x} \sigma_{A_{3}}^{z} \sigma_{B_{3}}^{x}$ \\
&& $\ket{01}$ & $\sigma_{A_{3}}^{x} \sigma_{A_{3}}^{z} \sigma_{A_{3}}^{x}\sigma_{B_{3}}^{x} $ \\
&& $\ket{10}$ & $\sigma_{A_{3}}^{x} \sigma_{A_{3}}^{z} $ \\
&& $\ket{11}$ & $\sigma_{A_{3}}^{x} \sigma_{A_{3}}^{z}  \sigma_{A_{3}}^{x}$\\
\cline{2-4}
&  \multirow{4}{*}{$\ket{\beta_{1}}_{A_{2}}\ket{\gamma_{0}}_{B_{2}}$} & $\ket{00}$ & $\sigma_{B_{3}}^{x}  \sigma_{B_{3}}^{z} \sigma_{B_{3}}^{x} $ \\
&& $\ket{01}$ & $ \sigma_{A_{3}}^{x} \sigma_{B_{3}}^{x} \sigma_{B_{3}}^{z} \sigma_{B_{3}}^{x}  $ \\
&& $\ket{10}$ & $ \sigma_{B_{3}}^{x} \sigma_{B_{3}}^{z}  $ \\
&& $\ket{11}$ & $ \sigma_{A_{3}}^{x} \sigma_{B_{3}}^{x} \sigma_{B_{3}}^{z}  $ \\
\cline{2-4}
&  \multirow{4}{*}{$\ket{\beta_{1}}_{A_{2}}\ket{\gamma_{1}}_{B_{2}}$} & $\ket{00}$ & $\sigma_{A_{3}}^{x} \sigma_{A_{3}}^{z}\sigma_{B_{3}}^{x} \sigma_{B_{3}}^{z} \sigma_{B_{3}}^{x}$ \\
&& $\ket{01}$ & $\sigma_{A_{3}}^{x} \sigma_{A_{3}}^{z} \sigma_{A_{3}}^{x} \sigma_{B_{3}}^{x} \sigma_{B_{3}}^{z} \sigma_{B_{3}}^{x}$ \\
&& $\ket{10}$ & $\sigma_{A_{3}}^{x} \sigma_{A_{3}}^{z} \sigma_{B_{3}}^{x} \sigma_{B_{3}}^{z} $ \\
&& $\ket{11}$ & $\sigma_{A_{3}}^{x} \sigma_{A_{3}}^{z} \sigma_{A_{3}}^{x} \sigma_{B_{3}}^{x} \sigma_{B_{3}}^{z} $ \\
[1.ex]
\hline
 \multirow{8}{*}{$\ket{\alpha_9}_{A_{1}B_{1}}$} &  \multirow{4}{*}{$\ket{\beta_{0}}_{A_{2}}\ket{\gamma_{0}}_{B_{2}}$} & $\ket{00}$ & $\sigma_{A_3}^z\sigma_{B_{3}}^{x} $ \\
&& $\ket{01}$ & $\sigma_{A_{3}}^{z}\sigma_{A_{3}}^{x}\sigma_{A_{3}}^{z} \sigma_{B_{3}}^{x} $ \\
&& $\ket{10} $ &  $\sigma_{A_3}^z$\\
&& $\ket{11} $ &  $\sigma_{A_{3}}^{z}\sigma_{A_{3}}^{x}\sigma_{A_{3}}^{z} $\\
\cline{2-4}
&  \multirow{4}{*}{$\ket{\beta_{0}}_{A_{2}}\ket{\gamma_{1}}_{B_{2}}$} & $\ket{00}$ & $\sigma_{A_{3}}^{x} \sigma_{B_{3}}^{x}$ \\
&& $\ket{01}$ & $ \sigma_{A_{3}}^{z} \sigma_{B_{3}}^{x} $ \\
&& $\ket{10}$ & $\sigma_{A_{3}}^{x} $ \\
&& $\ket{11}$ & $ \sigma_{A_{3}}^{z}  $\\
\cline{2-4}
&  \multirow{4}{*}{$\ket{\beta_{1}}_{A_{2}}\ket{\gamma_{0}}_{B_{2}}$} & $\ket{00}$ & $\sigma_{A_{3}}^{x}  \sigma_{A_{3}}^{z} \sigma_{A_{3}}^{x}\sigma_{B_{3}}^{z} $ \\
&& $\ket{01}$ & $ \sigma_{A_{3}}^{x}  \sigma_{B_{3}}^{z}  $ \\
&& $\ket{10}$ & $ \sigma_{A_{3}}^{x}  \sigma_{A_{3}}^{z} \sigma_{A_{3}}^{x}\sigma_{B_{3}}^{z} \sigma_{B_{3}}^{x} $ \\
&& $\ket{11}$ & $ \sigma_{A_{3}}^{x} \sigma_{B_{3}}^{z} \sigma_{B_{3}}^{x}  $ \\
\cline{2-4}
&  \multirow{4}{*}{$\ket{\beta_{1}}_{A_{2}}\ket{\gamma_{1}}_{B_{2}}$} & $\ket{00}$ & $\sigma_{A_{3}}^{x} \sigma_{B_{3}}^{x}\sigma_{B_{3}}^{z} \sigma_{B_{3}}^{x} $ \\
&& $\ket{01}$ & $\sigma_{A_{3}}^{z} \sigma_{B_{3}}^{x}\sigma_{B_{3}}^{z} \sigma_{B_{3}}^{x}$ \\
&& $\ket{10}$ & $\sigma_{A_{3}}^{x}  \sigma_{B_{3}}^{x} \sigma_{B_{3}}^{z} $ \\
&& $\ket{11}$ & $ \sigma_{A_{3}}^{z}  \sigma_{B_{3}}^{x} \sigma_{B_{3}}^{z} $ \\
\hline

\multirow{16}{*}{$\ket{\alpha_{10}}_{A_{1}B_{1}}$} & \multirow{4}{*}{$\ket{\beta_{0}}_{A_{2}}\ket{\gamma_{0}}_{B_{2}}$} & $\ket{00}$ & $\sigma_{B_{3}}^{x} $\\
&& $\ket{01}$ & $\sigma_{A_{3}}^{x} \sigma_{B_{3}}^{x} $ \\
&& $\ket{10} $ &  $\sigma_{B_{3}}^{z}$\\
&& $\ket{11} $ &  $ \sigma_{A_{3}}^{x} \sigma_{B_{3}}^{z}$\\
\cline{2-4}
& \multirow{4}{*}{$\ket{\beta_{0}}_{A_{2}}\ket{\gamma_{1}}_{B_{2}}$} & $\ket{00}$ & $\sigma_{A_{3}}^{x} \sigma_{A_{3}}^{z} \sigma_{B_{3}}^{x}$ \\
&& $\ket{01}$ & $\sigma_{A_{3}}^{x} \sigma_{A_{3}}^{z} \sigma_{A_{3}}^{x}\sigma_{B_{3}}^{x} $ \\
&& $\ket{10}$ & $\sigma_{A_{3}}^{x} \sigma_{A_{3}}^{z}\sigma_{B_{3}}^{z} $ \\
&& $\ket{11}$ & $\sigma_{A_{3}}^{x} \sigma_{A_{3}}^{z}  \sigma_{A_{3}}^{x}\sigma_{B_{3}}^{z}$\\
\cline{2-4}
& \multirow{4}{*}{$\ket{\beta_{1}}_{A_{2}}\ket{\gamma_{0}}_{B_{2}}$} & $\ket{00}$ & $\sigma_{B_{3}}^{x}  \sigma_{B_{3}}^{z} \sigma_{B_{3}}^{x} $ \\
&& $\ket{01}$ & $ \sigma_{A_{3}}^{x} \sigma_{B_{3}}^{x} \sigma_{B_{3}}^{z} \sigma_{B_{3}}^{x}  $ \\
&& $\ket{10}$ & $  \sigma_{B_{3}}^{x}  $ \\
&& $\ket{11}$ & $ \sigma_{A_{3}}^{x} \sigma_{B_{3}}^{x}   $ \\
\cline{2-4}
& \multirow{4}{*}{$\ket{\beta_{1}}_{A_{2}}\ket{\gamma_{1}}_{B_{2}}$} & $\ket{00}$ & $\sigma_{A_{3}}^{x}\sigma_{A_{3}}^{z} \sigma_{B_{3}}^{x}\sigma_{B_{3}}^{z} \sigma_{B_{3}}^{x}$ \\
&& $\ket{01}$ & $\sigma_{A_{3}}^{x} \sigma_{A_{3}}^{z} \sigma_{A_{3}}^{x}\sigma_{B_{3}}^{x} \sigma_{B_{3}}^{z} \sigma_{B_{3}}^{x}$ \\
&& $\ket{10}$ & $ \sigma_{A_{3}}^{x}\sigma_{A_{3}}^{z} \sigma_{B_{3}}^{x} $ \\
&& $\ket{11}$ & $\sigma_{A_{3}}^{x} \sigma_{A_{3}}^{z} \sigma_{A_{3}}^{x}  \sigma_{B_{3}}^{x} $ \\
[1.ex]
\hline
\multirow{16}{*}{$\ket{\alpha_{11}}_{A_{1}B_{1}}$} & \multirow{4}{*}{$\ket{\beta_{0}}_{A_{2}}\ket{\gamma_{0}}_{B_{2}}$} & $\ket{00}$ & $\sigma_{A_3}^z\sigma_{B_{3}}^{x} $ \\
&& $\ket{01}$ & $\sigma_{A_{3}}^{z}\sigma_{A_{3}}^{x}\sigma_{A_{3}}^{z} \sigma_{B_{3}}^{x} $ \\
&& $\ket{10} $ &  $\sigma_{A_3}^z\sigma_{B_3}^z$\\
&& $\ket{11} $ &  $\sigma_{A_{3}}^{z}\sigma_{A_{3}}^{x}\sigma_{A_{3}}^{z} \sigma_{B_{3}}^{z} $\\
\cline{2-4}
& \multirow{4}{*}{$\ket{\beta_{0}}_{A_{2}}\ket{\gamma_{1}}_{B_{2}}$} & $\ket{00}$ & $\sigma_{A_{3}}^{x} \sigma_{B_{3}}^{x}$ \\
&& $\ket{01}$ & $ \sigma_{A_{3}}^{z} \sigma_{B_{3}}^{x} $ \\
&& $\ket{10}$ & $\sigma_{A_{3}}^{x}\sigma_{B_{3}}^{z} $ \\
&& $\ket{11}$ & $ \sigma_{A_{3}}^{z} \sigma_{B_{3}}^{z} $\\
\cline{2-4}
& \multirow{4}{*}{$\ket{\beta_{1}}_{A_{2}}\ket{\gamma_{0}}_{B_{2}}$} & $\ket{00}$ & $  \sigma_{A_{3}}^{z} \sigma_{B_{3}}^{z} $ \\
&& $\ket{01}$ & $ \sigma_{A_{3}}^{x}  \sigma_{B_{3}}^{z}  $ \\
&& $\ket{10}$ & $ \sigma_{A_{3}}^{z}  \sigma_{B_{3}}^{x}  $ \\
&& $\ket{11}$ & $ \sigma_{A_{3}}^{x}  \sigma_{B_{3}}^{x}  $ \\
\cline{2-4}
& \multirow{4}{*}{$\ket{\beta_{1}}_{A_{2}}\ket{\gamma_{1}}_{B_{2}}$} & $\ket{00}$ & $\sigma_{A_{3}}^{x} \sigma_{B_{3}}^{x}\sigma_{B_{3}}^{z} \sigma_{B_{3}}^{x} $ \\
&& $\ket{01}$ & $\sigma_{A_{3}}^{z} \sigma_{B_{3}}^{x}\sigma_{B_{3}}^{z} \sigma_{B_{3}}^{x}$ \\
&& $\ket{10}$ & $\sigma_{A_{3}}^{x}  \sigma_{B_{3}}^{x} $ \\
&& $\ket{11}$ & $ \sigma_{A_{3}}^{z}  \sigma_{B_{3}}^{x}  $ \\
\hline

\multirow{16}{*}{$\ket{\alpha_{12}}_{A_{1}B_{1}}$} & \multirow{4}{*}{$\ket{\beta_{0}}_{A_{2}}\ket{\gamma_{0}}_{B_{2}}$} & $\ket{00}$ & $\sigma_{A_{3}}^{x} \sigma_{B_{3}}^{x} $\\
&& $\ket{01}$ & $  \sigma_{B_{3}}^{x} $ \\
&& $\ket{10} $ &  $\sigma_{A_{3}}^{x}$\\
&& $\ket{11} $ &  $ I $\\
\cline{2-4}
& \multirow{4}{*}{$\ket{\beta_{0}}_{A_{3}}\ket{\gamma_{1}}_{B_{3}}$} & $\ket{00}$ & $\sigma_{A_{3}}^{x} \sigma_{A_{3}}^{z} \sigma_{A_{3}}^{x}\sigma_{B_{3}}^{x}$ \\
&& $\ket{01}$ & $\sigma_{A_{3}}^{x} \sigma_{A_{3}}^{z} \sigma_{B_{3}}^{x} $ \\
&& $\ket{10}$ & $\sigma_{A_{3}}^{x} \sigma_{A_{3}}^{z}\sigma_{A_{3}}^{x} $ \\
&& $\ket{11}$ & $\sigma_{A_{3}}^{x} \sigma_{A_{3}}^{z}  $\\
\cline{2-4}
& \multirow{4}{*}{$\ket{\beta_{1}}_{A_{3}}\ket{\gamma_{0}}_{B_{3}}$} & $\ket{00}$ & $\sigma_{A_{3}}^{x} \sigma_{B_{3}}^{x}  \sigma_{B_{3}}^{z} \sigma_{B_{3}}^{x} $ \\
&& $\ket{01}$ & $  \sigma_{B_{3}}^{x} \sigma_{B_{3}}^{z} \sigma_{B_{3}}^{x}  $ \\
&& $\ket{10}$ & $ \sigma_{A_{3}}^{x} \sigma_{B_{3}}^{x} \sigma_{B_{3}}^{z}  $ \\
&& $\ket{11}$ & $  \sigma_{B_{3}}^{x} \sigma_{B_{3}}^{z}   $ \\
\cline{2-4}
& \multirow{4}{*}{$\ket{\beta_{1}}_{A_{3}}\ket{\gamma_{1}}_{B_{3}}$} & $\ket{00}$ & $\sigma_{A_{3}}^{x} \sigma_{A_{3}}^{z} \sigma_{A_{3}}^{x}\sigma_{B_{3}}^{x} \sigma_{B_{3}}^{z} \sigma_{B_{3}}^{x}$ \\
&& $\ket{01}$ & $\sigma_{A_{3}}^{x} \sigma_{A_{3}}^{z}  \sigma_{B_{3}}^{x} \sigma_{B_{3}}^{z} \sigma_{B_{3}}^{x}$ \\
&& $\ket{10}$ & $\sigma_{A_{3}}^{x} \sigma_{A_{3}}^{z} \sigma_{A_{3}}^{x} \sigma_{B_{3}}^{x} \sigma_{B_{3}}^{z} $ \\
&& $\ket{11}$ & $\sigma_{A_{3}}^{x} \sigma_{A_{3}}^{z}  \sigma_{B_{3}}^{x} \sigma_{B_{3}}^{z} $ \\
[1.ex]
\hline

\multirow{16}{*}{$\ket{\alpha_{13}}_{A_{1}B_{1}}$} & \multirow{4}{*}{$\ket{\beta_{0}}_{A_{2}}\ket{\gamma_{0}}_{B_{2}}$} & $\ket{00}$ & $\sigma_{A_{3}}^{x} \sigma_{B_{3}}^{x} $\\
&& $\ket{01}$ & $  \sigma_{A_{3}}^{z}\sigma_{B_{3}}^{x} $ \\
&& $\ket{10} $ &  $\sigma_{A_{3}}^{x}$\\
&& $\ket{11} $ &  $ \sigma_{A_{3}}^{z} $\\
\cline{2-4}
& \multirow{4}{*}{$\ket{\beta_{0}}_{A_{3}}\ket{\gamma_{1}}_{B_{3}}$} & $\ket{00}$ & $\sigma_{A_{3}}^{x} \sigma_{A_{3}}^{z} \sigma_{A_{3}}^{x}\sigma_{B_{3}}^{x}$ \\
&& $\ket{01}$ & $\sigma_{A_{3}}^{x}  \sigma_{B_{3}}^{x} $ \\
&& $\ket{10}$ & $\sigma_{A_{3}}^{x} \sigma_{A_{3}}^{z}\sigma_{A_{3}}^{x} $ \\
&& $\ket{11}$ & $\sigma_{A_{3}}^{x}   $\\
\cline{2-4}
& \multirow{4}{*}{$\ket{\beta_{1}}_{A_{3}}\ket{\gamma_{0}}_{B_{3}}$} & $\ket{00}$ & $\sigma_{A_{3}}^{x} \sigma_{B_{3}}^{x}  \sigma_{B_{3}}^{z} \sigma_{B_{3}}^{x} $ \\
&& $\ket{01}$ & $ \sigma_{A_{3}}^{z} \sigma_{B_{3}}^{x} \sigma_{B_{3}}^{z} \sigma_{B_{3}}^{x}  $ \\
&& $\ket{10}$ & $ \sigma_{A_{3}}^{x} \sigma_{B_{3}}^{x} \sigma_{B_{3}}^{z}  $ \\
&& $\ket{11}$ & $  \sigma_{A_{3}}^{z}\sigma_{B_{3}}^{x} \sigma_{B_{3}}^{z}   $ \\
\cline{2-4}
& \multirow{4}{*}{$\ket{\beta_{1}}_{A_{3}}\ket{\gamma_{1}}_{B_{3}}$} & $\ket{00}$ & $\sigma_{A_{3}}^{x} \sigma_{A_{3}}^{z} \sigma_{A_{3}}^{x}\sigma_{B_{3}}^{x} \sigma_{B_{3}}^{z} \sigma_{B_{3}}^{x}$ \\
&& $\ket{01}$ & $\sigma_{A_{3}}^{x}   \sigma_{B_{3}}^{x} \sigma_{B_{3}}^{z} \sigma_{B_{3}}^{x}$ \\
&& $\ket{10}$ & $\sigma_{A_{3}}^{x} \sigma_{A_{3}}^{z} \sigma_{A_{3}}^{x} \sigma_{B_{3}}^{x} \sigma_{B_{3}}^{z} $ \\
&& $\ket{11}$ & $\sigma_{A_{3}}^{x}  \sigma_{B_{3}}^{x} \sigma_{B_{3}}^{z} $ \\
\hline

\multirow{16}{*}{$\ket{\alpha_{14}}_{A_{1}B_{1}}$} & \multirow{4}{*}{$\ket{\beta_{0}}_{A_{2}}\ket{\gamma_{0}}_{B_{2}}$} & $\ket{00}$ & $\sigma_{A_{3}}^{x} \sigma_{B_{3}}^{x} $\\
&& $\ket{01}$ & $  \sigma_{B_{3}}^{x} $ \\
&& $\ket{10} $ &  $\sigma_{A_{3}}^{x}\sigma_{B_{3}}^{z}$\\
&& $\ket{11} $ &  $ \sigma_{B_{3}}^{z} $\\
\cline{2-4}
& \multirow{4}{*}{$\ket{\beta_{0}}_{A_{3}}\ket{\gamma_{1}}_{B_{3}}$} & $\ket{00}$ & $\sigma_{A_{3}}^{x} \sigma_{A_{3}}^{z} \sigma_{A_{3}}^{x}\sigma_{B_{3}}^{x}$ \\
&& $\ket{01}$ & $\sigma_{A_{3}}^{x} \sigma_{A_{3}}^{z} \sigma_{B_{3}}^{x} $ \\
&& $\ket{10}$ & $\sigma_{A_{3}}^{x} \sigma_{A_{3}}^{z}\sigma_{A_{3}}^{x}\sigma_{B_{3}}^{z} $ \\
&& $\ket{11}$ & $\sigma_{A_{3}}^{x} \sigma_{A_{3}}^{z}\sigma_{B_{3}}^{z}  $\\
\cline{2-4}
& \multirow{4}{*}{$\ket{\beta_{1}}_{A_{3}}\ket{\gamma_{0}}_{B_{3}}$} & $\ket{00}$ & $\sigma_{A_{3}}^{x} \sigma_{B_{3}}^{x}  \sigma_{B_{3}}^{z} \sigma_{B_{3}}^{x} $ \\
&& $\ket{01}$ & $  \sigma_{B_{3}}^{x} \sigma_{B_{3}}^{z} \sigma_{B_{3}}^{x}  $ \\
&& $\ket{10}$ & $ \sigma_{A_{3}}^{x} \sigma_{B_{3}}^{x}   $ \\
&& $\ket{11}$ & $  \sigma_{B_{3}}^{x}    $ \\
\cline{2-4}
& \multirow{4}{*}{$\ket{\beta_{1}}_{A_{3}}\ket{\gamma_{1}}_{B_{3}}$} & $\ket{00}$ & $\sigma_{A_{3}}^{x} \sigma_{A_{3}}^{z}\sigma_{A_{3}}^{x}\sigma_{B_{3}}^{x} \sigma_{B_{3}}^{z} \sigma_{B_{3}}^{x}$ \\
&& $\ket{01}$ & $  \sigma_{A_{3}}^{x} \sigma_{A_{3}}^{z}\sigma_{B_{3}}^{x} \sigma_{B_{3}}^{z} \sigma_{B_{3}}^{x}$ \\
&& $\ket{10}$ & $\sigma_{A_{3}}^{x} \sigma_{A_{3}}^{z} \sigma_{A_{3}}^{x} \sigma_{B_{3}}^{x}  $ \\
&& $\ket{11}$ & $\sigma_{A_{3}}^{x} \sigma_{A_{3}}^{z}  \sigma_{B_{3}}^{x}  $ \\
[1.ex]
\hline
\multirow{16}{*}{$\ket{\alpha_{15}}_{A_{1}B_{1}}$} & \multirow{4}{*}{$\ket{\beta_{0}}_{A_{2}}\ket{\gamma_{0}}_{B_{2}}$} & $\ket{00}$ & $\sigma_{A_{3}}^{x} \sigma_{B_{3}}^{x} $\\
&& $\ket{01}$ & $  \sigma_{A_{3}}^{z}\sigma_{B_{3}}^{x} $ \\
&& $\ket{10} $ &  $\sigma_{A_{3}}^{x}\sigma_{B_{3}}^{z}$\\
&& $\ket{11} $ &  $ \sigma_{A_{3}}^{z}\sigma_{B_{3}}^{z} $\\
\cline{2-4}
& \multirow{4}{*}{$\ket{\beta_{0}}_{A_{3}}\ket{\gamma_{1}}_{B_{3}}$} & $\ket{00}$ & $\sigma_{A_{3}}^{x} \sigma_{A_{3}}^{z}\sigma_{A_{3}}^{x}\sigma_{B_{3}}^{x}$ \\
&& $\ket{01}$ & $\sigma_{A_{3}}^{x}  \sigma_{B_{3}}^{x} $ \\
&& $\ket{10}$ & $\sigma_{A_{3}}^{x} \sigma_{A_{3}}^{z}\sigma_{A_{3}}^{x} \sigma_{B_{3}}^{z} $ \\
&& $\ket{11}$ & $\sigma_{A_{3}}^{x} \sigma_{B_{3}}^{z}  $\\
\cline{2-4}
& \multirow{4}{*}{$\ket{\beta_{1}}_{A_{3}}\ket{\gamma_{0}}_{B_{3}}$} & $\ket{00}$ & $\sigma_{A_{3}}^{x} \sigma_{B_{3}}^{x}  \sigma_{B_{3}}^{z} \sigma_{B_{3}}^{x} $ \\
&& $\ket{01}$ & $ \sigma_{A_{3}}^{z} \sigma_{B_{3}}^{x} \sigma_{B_{3}}^{z} \sigma_{B_{3}}^{x}  $ \\
&& $\ket{10}$ & $ \sigma_{A_{3}}^{x} \sigma_{B_{3}}^{x}  $ \\
&& $\ket{11}$ & $  \sigma_{A_{3}}^{z}\sigma_{B_{3}}^{x}   $ \\
\cline{2-4}
& \multirow{4}{*}{$\ket{\beta_{1}}_{A_{3}}\ket{\gamma_{1}}_{B_{3}}$} & $\ket{00}$ & $\sigma_{A_{3}}^{x} \sigma_{A_{3}}^{z} \sigma_{A_{3}}^{x}\sigma_{B_{3}}^{x} \sigma_{B_{3}}^{z} \sigma_{B_{3}}^{x}$ \\
&& $\ket{01}$ & $\sigma_{A_{3}}^{x}   \sigma_{B_{3}}^{x} \sigma_{B_{3}}^{z} \sigma_{B_{3}}^{x}$ \\
&& $\ket{10}$ & $\sigma_{A_{3}}^{x} \sigma_{A_{3}}^{z} \sigma_{A_{3}}^{x} \sigma_{B_{3}}^{x}  $ \\
&& $\ket{11}$ & $\sigma_{A_{3}}^{x}  \sigma_{B_{3}}^{x}  $ \\
\hline
\label{table2}
\end{supertabular}
\end{center}
\FloatBarrier

\section{Complete table for Controlled Bidirectional Remote State Preparation protocol on two-vertex complete graph}
The controlled bidirectional remote state preparation protocol on a two-vertex complete graph is summarized in the following table.
\begin{center}
\captionof{table}{Outcome of measurement performed by Alice and Bob in the first coin state and the unitary operation on their second coin state and measurement performed by  Charlie in his two coin state for creating the known single qubit state.}
\begin{supertabular}{ p{3cm} p{4cm} p{3.5cm} p{4.5cm} }
\hline
Position basis~~~ &  $A_{2},B_{2}$ Coin basis~~~~ & $C_{1},C_{2}$ Coin basis  ~~~&  Unitary operation ~~~~~~ \\
\hline

\multirow{16}{*}{$\ket{\alpha_0}_{A_{1}B_{1}}$} & \multirow{4}{*}{$\ket{\beta_{0}}_{A_{2}}\ket{\gamma_{0}}_{B_{2}}$} & $\ket{00}$ & I \\
&& $\ket{01} $ &  $\sigma_{B_{3}}^{x}$\\
&& $\ket{10} $ &  $\sigma_{A_{3}}^{x}$\\
&& $\ket{11} $ &  $\sigma_{A_{3}}^{x} \sigma_{B_{3}}^{x} $ \\
\cline{2-4}
& \multirow{4}{*}{$\ket{\beta_{0}}_{A_{2}}\ket{\gamma_{1}}_{B_{2}}$} & $\ket{00}$ & $\sigma_{A_{3}}^{x} \sigma_{A_{3}}^{z}$ \\
&& $\ket{01} $ &  $\sigma_{A_{3}}^{x}\sigma_{A_{3}}^{z} \sigma_{A_{3}}^{x}$\\
&& $\ket{10} $ &  $\sigma_{A_{3}}^{x}\sigma_{A_{3}}^{z} \sigma_{B_{3}}^{x}$\\
&& $\ket{11} $ &  $\sigma_{A_{3}}^{x} \sigma_{A_{3}}^{z} \sigma_{A_{3}}^{x} \sigma_{B_{3}}^{x}$ \\
\cline{2-4}
& \multirow{4}{*}{$\ket{\beta_{1}}_{A_{2}}\ket{\gamma_{0}}_{B_{2}}$} & $\ket{00}$ & $\sigma_{B_{3}}^{x} \sigma_{B_{3}}^{z}$ \\
&& $\ket{01} $ &  $\sigma_{A_{3}}^{x} \sigma_{B_{3}}^{x} \sigma_{B_{3}}^{z} $\\
&& $\ket{10} $ &  $ \sigma_{B_{3}}^{x} \sigma_{B_{3}}^{z} \sigma_{B_{3}}^{x} $\\
&& $\ket{11} $ &  $\sigma_{A_{3}}^{x} \sigma_{B_{3}}^{x} \sigma_{B_{3}}^{z} \sigma_{B_{3}}^{x}$\\
\cline{2-4}
& \multirow{4}{*}{$\ket{\beta_{1}}_{A_{2}}\ket{\gamma_{1}}_{B_{2}}$} & $\ket{00}$ & $\sigma_{A_{3}}^{x} \sigma_{A_{3}}^{z} \sigma_{B_{3}}^{x} \sigma_{B_{3}}^{z}$ \\
&& $\ket{01}$ & $\sigma_{A_{3}}^{x} \sigma_{A_{3}}^{z} \sigma_{A_{3}}^{x} \sigma_{B_{3}}^{x} \sigma_{B_{3}}^{z}$ \\
&& $\ket{10}$ & $\sigma_{A_{3}}^{x} \sigma_{A_{3}}^{z}  \sigma_{B_{3}}^{x} \sigma_{B_{3}}^{z} \sigma_{B_{3}}^{x}$ \\
&& $\ket{11}$ & $\sigma_{A_{3}}^{x} \sigma_{A_{3}}^{z} \sigma_{A_{3}}^{x} \sigma_{B_{3}}^{x} \sigma_{B_{3}}^{z} \sigma_{B_{3}}^{x}$ \\
[1.ex]
\hline

\multirow{16}{*}{$\ket{\alpha_1}_{A_{1}B_{1}}$} & \multirow{4}{*}{$\ket{\beta_{0}}_{A_{2}}\ket{\gamma_{0}}_{B_{2}}$} & $\ket{00}$ & $\sigma_{A_{3}}^{x} $\\
&& $\ket{01}$ & I \\
&& $\ket{10} $ &  $\sigma_{A_{3}}^{x} \sigma_{B_{3}}^{x} $\\
&& $\ket{11} $ &  $ \sigma_{B_{3}}^{x} $\\
\cline{2-4}
& \multirow{4}{*}{$\ket{\beta_{0}}_{A_{2}}\ket{\gamma_{1}}_{B_{2}}$} & $\ket{00}$ & $\sigma_{A_{3}}^{x} \sigma_{A_{3}}^{z} \sigma_{A_{3}}^{x}$ \\
&& $\ket{01}$ & $\sigma_{A_{3}}^{x} \sigma_{A_{3}}^{z} $ \\
&& $\ket{10}$ & $\sigma_{A_{3}}^{x} \sigma_{A_{3}}^{z} \sigma_{A_{3}}^{x} \sigma_{B_{3}}^{x}$ \\
&& $\ket{11}$ & $\sigma_{A_{3}}^{x} \sigma_{A_{3}}^{z}  \sigma_{B_{3}}^{x}$\\
\cline{2-4}
& \multirow{4}{*}{$\ket{\beta_{1}}_{A_{2}}\ket{\gamma_{0}}_{B_{2}}$} & $\ket{00}$ & $\sigma_{A_{3}}^{x}  \sigma_{B_{3}}^{x} \sigma_{B_{3}}^{z}$ \\
&& $\ket{01}$ & $\sigma_{B_{3}}^{x} \sigma_{B_{3}}^{z} $ \\
&& $\ket{10}$ & $ \sigma_{A_{3}}^{x}\sigma_{B_{3}}^{x} \sigma_{B_{3}}^{z} \sigma_{B_{3}}^{x} $ \\
&& $\ket{11}$ & $ \sigma_{B_{3}}^{x} \sigma_{B_{3}}^{z} \sigma_{B_{3}}^{x} $ \\
\cline{2-4}
& \multirow{4}{*}{$\ket{\beta_{1}}_{A_{2}}\ket{\gamma_{1}}_{B_{2}}$} & $\ket{00}$ & $\sigma_{A_{3}}^{x} \sigma_{A_{3}}^{z}\sigma_{A_{3}}^{x} \sigma_{B_{3}}^{x} \sigma_{B_{3}}^{z}$ \\
&& $\ket{01}$ & $\sigma_{A_{3}}^{x} \sigma_{A_{3}}^{z} \sigma_{B_{3}}^{x} \sigma_{B_{3}}^{z}$ \\
&& $\ket{10}$ & $\sigma_{A_{3}}^{x} \sigma_{A_{3}}^{z}\sigma_{A_{3}}^{x} \sigma_{B_{3}}^{x} \sigma_{B_{3}}^{z} \sigma_{B_{3}}^{x}$ \\
&& $\ket{11}$ & $\sigma_{A_{3}}^{x} \sigma_{A_{3}}^{z} \sigma_{B_{3}}^{x} \sigma_{B_{3}}^{z} \sigma_{B_{3}}^{x}$ \\
[1.ex]
\hline

\multirow{16}{*}{$\ket{\alpha_2}_{A_{1}B_{1}}$} & \multirow{4}{*}{$\ket{\beta_{0}}_{A_{2}}\ket{\gamma_{0}}_{B_{2}}$} & $\ket{00}$ & $\sigma_{B_{3}}^{x} $\\
&& $\ket{01}$ & $\sigma_{A_{3}}^{x} \sigma_{B_{3}}^{x} $ \\
&& $\ket{10} $ &  I\\
&& $\ket{11} $ &  $ \sigma_{A_{3}}^{x} $\\
\cline{2-4}
& \multirow{4}{*}{$\ket{\beta_{0}}_{A_{2}}\ket{\gamma_{1}}_{B_{2}}$} & $\ket{00}$ & $\sigma_{A_{3}}^{x} \sigma_{A_{3}}^{z} \sigma_{B_{3}}^{x}$ \\
&& $\ket{01}$ & $\sigma_{A_{3}}^{x} \sigma_{A_{3}}^{z} \sigma_{A_{3}}^{x}\sigma_{B_{3}}^{x} $ \\
&& $\ket{10}$ & $\sigma_{A_{3}}^{x} \sigma_{A_{3}}^{z} $ \\
&& $\ket{11}$ & $\sigma_{A_{3}}^{x} \sigma_{A_{3}}^{z}  \sigma_{A_{3}}^{x}$\\
\cline{2-4}
& \multirow{4}{*}{$\ket{\beta_{1}}_{A_{2}}\ket{\gamma_{0}}_{B_{2}}$} & $\ket{00}$ & $\sigma_{B_{3}}^{x}  \sigma_{B_{3}}^{z} \sigma_{B_{3}}^{x} $ \\
&& $\ket{01}$ & $ \sigma_{A_{3}}^{x} \sigma_{B_{3}}^{x} \sigma_{B_{3}}^{z} \sigma_{B_{3}}^{x}  $ \\
&& $\ket{10}$ & $ \sigma_{B_{3}}^{x} \sigma_{B_{3}}^{z}  $ \\
&& $\ket{11}$ & $ \sigma_{A_{3}}^{x} \sigma_{B_{3}}^{x} \sigma_{B_{3}}^{z}  $ \\
\cline{2-4}
& \multirow{4}{*}{$\ket{\beta_{1}}_{A_{2}}\ket{\gamma_{1}}_{B_{2}}$} & $\ket{00}$ & $\sigma_{A_{3}}^{x} \sigma_{A_{3}}^{z}\sigma_{B_{3}}^{x} \sigma_{B_{3}}^{z} \sigma_{B_{3}}^{x}$ \\
&& $\ket{01}$ & $\sigma_{A_{3}}^{x} \sigma_{A_{3}}^{z} \sigma_{A_{3}}^{x} \sigma_{B_{3}}^{x} \sigma_{B_{3}}^{z} \sigma_{B_{3}}^{x}$ \\
&& $\ket{10}$ & $\sigma_{A_{3}}^{x} \sigma_{A_{3}}^{z} \sigma_{B_{3}}^{x} \sigma_{B_{3}}^{z} $ \\
&& $\ket{11}$ & $\sigma_{A_{3}}^{x} \sigma_{A_{3}}^{z} \sigma_{A_{3}}^{x} \sigma_{B_{3}}^{x} \sigma_{B_{3}}^{z} $ \\
[1.ex]
\hline

\multirow{16}{*}{$\ket{\alpha_3}_{A_{1}B_{1}}$} & \multirow{4}{*}{$\ket{\beta_{0}}_{A_{2}}\ket{\gamma_{0}}_{B_{2}}$} & $\ket{00}$ & $\sigma_{A_{3}}^{x} \sigma_{B_{3}}^{x} $\\
&& $\ket{01}$ & $  \sigma_{B_{3}}^{x} $ \\
&& $\ket{10} $ &  $\sigma_{A_{3}}^{x}$\\
&& $\ket{11} $ &  $ I $\\
\cline{2-4}
& \multirow{4}{*}{$\ket{\beta_{0}}_{A_{3}}\ket{\gamma_{1}}_{B_{3}}$} & $\ket{00}$ & $\sigma_{A_{3}}^{x} \sigma_{A_{3}}^{z} \sigma_{A_{3}}^{x}\sigma_{B_{3}}^{x}$ \\
&& $\ket{01}$ & $\sigma_{A_{3}}^{x} \sigma_{A_{3}}^{z} \sigma_{B_{3}}^{x} $ \\
&& $\ket{10}$ & $\sigma_{A_{3}}^{x} \sigma_{A_{3}}^{z}\sigma_{A_{3}}^{x} $ \\
&& $\ket{11}$ & $\sigma_{A_{3}}^{x} \sigma_{A_{3}}^{z}  $\\
\cline{2-4}
& \multirow{4}{*}{$\ket{\beta_{1}}_{A_{3}}\ket{\gamma_{0}}_{B_{3}}$} & $\ket{00}$ & $\sigma_{A_{3}}^{x} \sigma_{B_{3}}^{x}  \sigma_{B_{3}}^{z} \sigma_{B_{3}}^{x} $ \\
&& $\ket{01}$ & $  \sigma_{B_{3}}^{x} \sigma_{B_{3}}^{z} \sigma_{B_{3}}^{x}  $ \\
&& $\ket{10}$ & $ \sigma_{A_{3}}^{x} \sigma_{B_{3}}^{x} \sigma_{B_{3}}^{z}  $ \\
&& $\ket{11}$ & $  \sigma_{B_{3}}^{x} \sigma_{B_{3}}^{z}   $ \\
\cline{2-4}
& \multirow{4}{*}{$\ket{\beta_{1}}_{A_{3}}\ket{\gamma_{1}}_{B_{3}}$} & $\ket{00}$ & $\sigma_{A_{3}}^{x} \sigma_{A_{3}}^{z} \sigma_{A_{3}}^{x}\sigma_{B_{3}}^{x} \sigma_{B_{3}}^{z} \sigma_{B_{3}}^{x}$ \\
&& $\ket{01}$ & $\sigma_{A_{3}}^{x} \sigma_{A_{3}}^{z}  \sigma_{B_{3}}^{x} \sigma_{B_{3}}^{z} \sigma_{B_{3}}^{x}$ \\
&& $\ket{10}$ & $\sigma_{A_{3}}^{x} \sigma_{A_{3}}^{z} \sigma_{A_{3}}^{x} \sigma_{B_{3}}^{x} \sigma_{B_{3}}^{z} $ \\
&& $\ket{11}$ & $\sigma_{A_{3}}^{x} \sigma_{A_{3}}^{z}  \sigma_{B_{3}}^{x} \sigma_{B_{3}}^{z} $ \\
\hline
\label{table3}
\end{supertabular}
\end{center}
\FloatBarrier

\noindent
\bibliographystyle{unsrt}
\bibliography{ref.bib}

@article{BEN1993,
  title = {Teleporting an unknown quantum state via dual classical and Einstein-Podolsky-Rosen channels},
  author = {Bennett, Charles H. and Brassard, Gilles and Cr\'epeau, Claude and Jozsa, Richard and Peres, Asher and Wootters, William K.},
  journal = {Phys. Rev. Lett.},
  volume = {70},
  issue = {13},
  pages = {1895--1899},
  numpages = {0},
  year = {1993},
  month = {Mar},
  publisher = {American Physical Society},
  doi = {10.1103/PhysRevLett.70.1895},
  url = {https://link.aps.org/doi/10.1103/PhysRevLett.70.1895}
}

@Article{BOU1997,
author={Bouwmeester, Dik
and Pan, Jian-Wei
and Mattle, Klaus
and Eibl, Manfred
and Weinfurter, Harald
and Zeilinger, Anton},
title={Experimental quantum teleportation},
journal={Nature},
year={1997},
month={Dec},
day={01},
volume={390},
number={6660},
pages={575-579},
issn={1476-4687},
doi={10.1038/37539},
url={https://doi.org/10.1038/37539}
}

@article{RIG2005,
  title = {Quantum teleportation of an arbitrary two-qubit state and its relation to multipartite entanglement},
  author = {Rigolin, Gustavo},
  journal = {Phys. Rev. A},
  volume = {71},
  issue = {3},
  pages = {032303},
  numpages = {5},
  year = {2005},
  month = {Mar},
  publisher = {American Physical Society},
  doi = {10.1103/PhysRevA.71.032303},
  url = {https://link.aps.org/doi/10.1103/PhysRevA.71.032303}
}

@article{BEN1992,
  title = {Communication via one- and two-particle operators on Einstein-Podolsky-Rosen states},
  author = {Bennett, Charles H. and Wiesner, Stephen J.},
  journal = {Phys. Rev. Lett.},
  volume = {69},
  issue = {20},
  pages = {2881--2884},
  numpages = {0},
  year = {1992},
  month = {Nov},
  publisher = {American Physical Society},
  doi = {10.1103/PhysRevLett.69.2881},
  url = {https://link.aps.org/doi/10.1103/PhysRevLett.69.2881}
}

@Article{SEB2023,
author={Sebastian, Arun
and Mansar, Afnan. N.
and Randeep, N. C.},
title={Beyond Qubits: An Extensive Noise Analysis for Qutrit Quantum Teleportation},
journal={International Journal of Theoretical Physics},
year={2023},
month={Nov},
day={30},
volume={62},
number={12},
pages={258},
issn={1572-9575},
doi={10.1007/s10773-023-05515-5},
url={https://doi.org/10.1007/s10773-023-05515-5}
}

@Article{RAN2024,
author={Randeep, N. C.
and Anukrishna, C.
and Neha Raj, A. K.},
title={Quantum teleportation scheme using entangled two ququads and its noise effects},
journal={Quantum Information Processing},
year={2024},
month={May},
day={14},
volume={23},
number={5},
pages={192},
issn={1573-1332},
doi={10.1007/s11128-024-04381-2},
url={https://doi.org/10.1007/s11128-024-04381-2}
}

@article{LO2000,
  title = {Classical-communication cost in distributed quantum-information processing: A generalization of quantum-communication complexity},
  author = {Lo, Hoi-Kwong},
  journal = {Phys. Rev. A},
  volume = {62},
  issue = {1},
  pages = {012313},
  numpages = {7},
  year = {2000},
  month = {Jun},
  publisher = {American Physical Society},
  doi = {10.1103/PhysRevA.62.012313},
  url = {https://link.aps.org/doi/10.1103/PhysRevA.62.012313}
}

@article{BEN2001,
  title = {Remote State Preparation},
  author = {Bennett, Charles H. and DiVincenzo, David P. and Shor, Peter W. and Smolin, John A. and Terhal, Barbara M. and Wootters, William K.},
  journal = {Phys. Rev. Lett.},
  volume = {87},
  issue = {7},
  pages = {077902},
  numpages = {4},
  year = {2001},
  month = {Jul},
  publisher = {American Physical Society},
  doi = {10.1103/PhysRevLett.87.077902},
  url = {https://link.aps.org/doi/10.1103/PhysRevLett.87.077902}
}

@article{PET2005,
  title = {Remote State Preparation: Arbitrary Remote Control of Photon Polarization},
  author = {Peters, Nicholas A. and Barreiro, Julio T. and Goggin, Michael E. and Wei, Tzu-Chieh and Kwiat, Paul G.},
  journal = {Phys. Rev. Lett.},
  volume = {94},
  issue = {15},
  pages = {150502},
  numpages = {4},
  year = {2005},
  month = {Apr},
  publisher = {American Physical Society},
  doi = {10.1103/PhysRevLett.94.150502},
  url = {https://link.aps.org/doi/10.1103/PhysRevLett.94.150502}
}

@article{DEV2001,
  title = {Low-Entanglement Remote State Preparation},
  author = {Devetak, Igor and Berger, Toby},
  journal = {Phys. Rev. Lett.},
  volume = {87},
  issue = {19},
  pages = {197901},
  numpages = {4},
  year = {2001},
  month = {Oct},
  publisher = {American Physical Society},
  doi = {10.1103/PhysRevLett.87.197901},
  url = {https://link.aps.org/doi/10.1103/PhysRevLett.87.197901}
}

@article{BER2003,
  title = {Optimal Remote State Preparation},
  author = {Berry, Dominic W. and Sanders, Barry C.},
  journal = {Phys. Rev. Lett.},
  volume = {90},
  issue = {5},
  pages = {057901},
  numpages = {4},
  year = {2003},
  month = {Feb},
  publisher = {American Physical Society},
  doi = {10.1103/PhysRevLett.90.057901},
  url = {https://link.aps.org/doi/10.1103/PhysRevLett.90.057901}
}

@article{EKE1991,
  title = {Quantum cryptography based on Bell's theorem},
  author = {Ekert, Artur K.},
  journal = {Phys. Rev. Lett.},
  volume = {67},
  issue = {6},
  pages = {661--663},
  numpages = {0},
  year = {1991},
  month = {Aug},
  publisher = {American Physical Society},
  doi = {10.1103/PhysRevLett.67.661},
  url = {https://link.aps.org/doi/10.1103/PhysRevLett.67.661}
}

@article{PIR2020,
author = {S. Pirandola and U. L. Andersen and L. Banchi and M. Berta and D. Bunandar and R. Colbeck and D. Englund and T. Gehring and C. Lupo and C. Ottaviani and J. L. Pereira and M. Razavi and J. Shamsul Shaari and M. Tomamichel and V. C. Usenko and G. Vallone and P. Villoresi and P. Wallden},
journal = {Adv. Opt. Photon.},
number = {4},
pages = {1012--1236},
publisher = {Optica Publishing Group},
title = {Advances in quantum cryptography},
volume = {12},
month = {Dec},
year = {2020},
url = {https://opg.optica.org/aop/abstract.cfm?URI=aop-12-4-1012},
doi = {10.1364/AOP.361502},
}

@article{GIS2002,
  title = {Quantum cryptography},
  author = {Gisin, Nicolas and Ribordy, Gr\'egoire and Tittel, Wolfgang and Zbinden, Hugo},
  journal = {Rev. Mod. Phys.},
  volume = {74},
  issue = {1},
  pages = {145--195},
  numpages = {0},
  year = {2002},
  month = {Mar},
  publisher = {American Physical Society},
  doi = {10.1103/RevModPhys.74.145},
  url = {https://link.aps.org/doi/10.1103/RevModPhys.74.145}
}

@article{HAR2004,
  title = {Superdense Coding of Quantum States},
  author = {Harrow, Aram and Hayden, Patrick and Leung, Debbie},
  journal = {Phys. Rev. Lett.},
  volume = {92},
  issue = {18},
  pages = {187901},
  numpages = {4},
  year = {2004},
  month = {May},
  publisher = {American Physical Society},
  doi = {10.1103/PhysRevLett.92.187901},
  url = {https://link.aps.org/doi/10.1103/PhysRevLett.92.187901}
}

@article{SHA2012,
  title = {Distributed superdense coding over noisy channels},
  author = {Shadman, Z. and Kampermann, H. and Bru\ss{}, D. and Macchiavello, C.},
  journal = {Phys. Rev. A},
  volume = {85},
  issue = {5},
  pages = {052306},
  numpages = {9},
  year = {2012},
  month = {May},
  publisher = {American Physical Society},
  doi = {10.1103/PhysRevA.85.052306},
  url = {https://link.aps.org/doi/10.1103/PhysRevA.85.052306}
}

@article{HIL1999,
  title = {Quantum secret sharing},
  author = {Hillery, Mark and Bu\ifmmode \check{z}\else \v{z}\fi{}ek, Vladim\'{\i}r and Berthiaume, Andr\'e},
  journal = {Phys. Rev. A},
  volume = {59},
  issue = {3},
  pages = {1829--1834},
  numpages = {0},
  year = {1999},
  month = {Mar},
  publisher = {American Physical Society},
  doi = {10.1103/PhysRevA.59.1829},
  url = {https://link.aps.org/doi/10.1103/PhysRevA.59.1829}
}

@article{GOT2000,
  title = {Theory of quantum secret sharing},
  author = {Gottesman, Daniel},
  journal = {Phys. Rev. A},
  volume = {61},
  issue = {4},
  pages = {042311},
  numpages = {8},
  year = {2000},
  month = {Mar},
  publisher = {American Physical Society},
  doi = {10.1103/PhysRevA.61.042311},
  url = {https://link.aps.org/doi/10.1103/PhysRevA.61.042311}
}

@article{TIT2001,
  title = {Experimental demonstration of quantum secret sharing},
  author = {Tittel, W. and Zbinden, H. and Gisin, N.},
  journal = {Phys. Rev. A},
  volume = {63},
  issue = {4},
  pages = {042301},
  numpages = {6},
  year = {2001},
  month = {Mar},
  publisher = {American Physical Society},
  doi = {10.1103/PhysRevA.63.042301},
  url = {https://link.aps.org/doi/10.1103/PhysRevA.63.042301}
}

@article{SIN2005,
  title = {Generalized quantum secret sharing},
  author = {Singh, Sudhir Kumar and Srikanth, R.},
  journal = {Phys. Rev. A},
  volume = {71},
  issue = {1},
  pages = {012328},
  numpages = {6},
  year = {2005},
  month = {Jan},
  publisher = {American Physical Society},
  doi = {10.1103/PhysRevA.71.012328},
  url = {https://link.aps.org/doi/10.1103/PhysRevA.71.012328}
}

@article{NGU2004,
title = {Quantum dialogue},
journal = {Physics Letters A},
volume = {328},
number = {1},
pages = {6-10},
year = {2004},
issn = {0375-9601},
doi = {https://doi.org/10.1016/j.physleta.2004.06.009},
url = {https://www.sciencedirect.com/science/article/pii/S0375960104007868},
author = {Ba An Nguyen}
}

@Article{MAN2005,
title = {Quantum Dialogue Revisited},
journal = {Chin. Phys. Lett.},
volume = {22},
number = {1},
pages = {22-24},
year = {2005},
issn = {},
	
url = {http://cpl.iphy.ac.cn/en/article/id/37306},
author = {MAN Zhong-Xiao and ZHANG Zhan-Jun and LI Yong}
}

@article{GAO2010,
title = {Two quantum dialogue protocols without information leakage},
journal = {Optics Communications},
volume = {283},
number = {10},
pages = {2288-2293},
year = {2010},
issn = {0030-4018},
doi = {https://doi.org/10.1016/j.optcom.2010.01.022},
url = {https://www.sciencedirect.com/science/article/pii/S003040181000026X},
author = {Gan Gao}
}

@article{ABA2006,
  title = {Quantum walk on the line: Entanglement and nonlocal initial conditions},
  author = {Abal, G. and Siri, R. and Romanelli, A. and Donangelo, R.},
  journal = {Phys. Rev. A},
  volume = {73},
  issue = {4},
  pages = {042302},
  numpages = {9},
  year = {2006},
  month = {Apr},
  publisher = {American Physical Society},
  doi = {10.1103/PhysRevA.73.042302},
  url = {https://link.aps.org/doi/10.1103/PhysRevA.73.042302}
}

@article{CAR2005,
doi = {10.1088/1367-2630/7/1/156},
url = {https://doi.org/10.1088/1367-2630/7/1/156},
year = {2005},
month = {jul},
publisher = {},
volume = {7},
number = {1},
pages = {156},
author = {Carneiro, Ivens and Loo, Meng and Xu, Xibai and Girerd, Mathieu and Kendon, Viv and Knight, Peter L},
title = {Entanglement in coined quantum walks on regular graphs},
journal = {New Journal of Physics}
}

@Article{VEN2012,
author={Venegas-Andraca, Salvador El{\'i}as},
title={Quantum walks: a comprehensive review},
journal={Quantum Information Processing},
year={2012},
month={Oct},
day={01},
volume={11},
number={5},
pages={1015-1106},
issn={1573-1332},
doi={10.1007/s11128-012-0432-5},
url={https://doi.org/10.1007/s11128-012-0432-5}
}

@article{KEM2003,
author = {J Kempe},
title = {Quantum random walks: An introductory overview},
journal = {Contemporary Physics},
volume = {44},
number = {4},
pages = {307--327},
year = {2003},
publisher = {Taylor \& Francis},
doi = {10.1080/00107151031000110776},
URL = { https://doi.org/10.1080/00107151031000110776
    },
eprint = { https://doi.org/10.1080/00107151031000110776}
}

@article{KAD2021,
title = {Quantum walk and its application domains: A systematic review},
journal = {Computer Science Review},
volume = {41},
pages = {100419},
year = {2021},
issn = {1574-0137},
doi = {https://doi.org/10.1016/j.cosrev.2021.100419},
url = {https://www.sciencedirect.com/science/article/pii/S1574013721000599},
author = {Karuna Kadian and Sunita Garhwal and Ajay Kumar}
}

@article{ZAR2010,
  title = {Realization of a Quantum Walk with One and Two Trapped Ions},
  author = {Z\"ahringer, F. and Kirchmair, G. and Gerritsma, R. and Solano, E. and Blatt, R. and Roos, C. F.},
  journal = {Phys. Rev. Lett.},
  volume = {104},
  issue = {10},
  pages = {100503},
  numpages = {4},
  year = {2010},
  month = {Mar},
  publisher = {American Physical Society},
  doi = {10.1103/PhysRevLett.104.100503},
  url = {https://link.aps.org/doi/10.1103/PhysRevLett.104.100503}
}

@article{SCH2010,
  title = {Photons Walking the Line: A Quantum Walk with Adjustable Coin Operations},
  author = {Schreiber, A. and Cassemiro, K. N. and Poto\ifmmode \check{c}\else \v{c}\fi{}ek, V. and G\'abris, A. and Mosley, P. J. and Andersson, E. and Jex, I. and Silberhorn, Ch.},
  journal = {Phys. Rev. Lett.},
  volume = {104},
  issue = {5},
  pages = {050502},
  numpages = {4},
  year = {2010},
  month = {Feb},
  publisher = {American Physical Society},
  doi = {10.1103/PhysRevLett.104.050502},
  url = {https://link.aps.org/doi/10.1103/PhysRevLett.104.050502}
}

@article{RYA2005,
  title = {Experimental implementation of a discrete-time quantum random walk on an NMR quantum-information processor},
  author = {Ryan, C. A. and Laforest, M. and Boileau, J. C. and Laflamme, R.},
  journal = {Phys. Rev. A},
  volume = {72},
  issue = {6},
  pages = {062317},
  numpages = {8},
  year = {2005},
  month = {Dec},
  publisher = {American Physical Society},
  doi = {10.1103/PhysRevA.72.062317},
  url = {https://link.aps.org/doi/10.1103/PhysRevA.72.062317}
}

@Article{JIA2021,
author={Jia-yin, Peng
and Hong-xuan, Lei},
title={Cyclic Remote State Preparation},
journal={International Journal of Theoretical Physics},
year={2021},
month={Apr},
day={01},
volume={60},
number={4},
pages={1593-1602},
issn={1572-9575},
doi={10.1007/s10773-021-04782-4},
url={https://doi.org/10.1007/s10773-021-04782-4}
}

@Article{WAN2017,
author={Wang, Yu
and Shang, Yun
and Xue, Peng},
title={Generalized teleportation by quantum walks},
journal={Quantum Information Processing},
year={2017},
month={Jul},
day={26},
volume={16},
number={9},
pages={221},
issn={1573-1332},
doi={10.1007/s11128-017-1675-y},
url={https://doi.org/10.1007/s11128-017-1675-y}
}

@Article{LI2019,
author={Li, Heng-Ji
and Chen, Xiu-Bo
and Wang, Ya-Lan
and Hou, Yan-Yan
and Li, Jian},
title={A new kind of flexible quantum teleportation of an arbitrary multi-qubit state by multi-walker quantum walks},
journal={Quantum Information Processing},
year={2019},
month={Jul},
day={17},
volume={18},
number={9},
pages={266},
issn={1573-1332},
doi={10.1007/s11128-019-2374-7},
url={https://doi.org/10.1007/s11128-019-2374-7}
}

@Article{ZAR2023,
author={Zarmehi, Fahimeh
and Talebi, Siamak
and Pourkarimi, Mohammad Reza},
title={Quantum walk-based controlled quantum teleportation schemes under the effect of decoherence in Markovian and non-Markovian regimes},
journal={Quantum Information Processing},
year={2023},
month={Dec},
day={12},
volume={22},
number={12},
pages={436},
issn={1573-1332},
doi={10.1007/s11128-023-04190-z},
url={https://doi.org/10.1007/s11128-023-04190-z}
}

@Article{SHI2022,
author={Shi, Wei-Min
and Bai, Meng-Xuan
and Zhou, Yi-Hua
and Yang, Yu-Guang},
title={Controlled quantum teleportation based on quantum walks},
journal={Quantum Information Processing},
year={2022},
month={Dec},
day={21},
volume={22},
number={1},
pages={34},
issn={1573-1332},
doi={10.1007/s11128-022-03737-w},
url={https://doi.org/10.1007/s11128-022-03737-w}
}

@article{SHA2018,
doi = {10.1209/0295-5075/124/60009},
url = {https://doi.org/10.1209/0295-5075/124/60009},
year = {2019},
month = {jan},
publisher = {EDP Sciences, IOP Publishing and Società Italiana di Fisica},
volume = {124},
number = {6},
pages = {60009},
author = {Shang, Yun and Wang, Yu and Li, Meng and Lu, Ruqian},
title = {Quantum communication protocols by quantum walks with two coins},
journal = {Europhysics Letters}}

@Article{CHA2019,
author={Chatterjee, Yagnik
and Devrari, Vipin
and Behera, Bikash K.
and Panigrahi, Prasanta K.},
title={Experimental realization of quantum teleportation using coined quantum walks},
journal={Quantum Information Processing},
year={2019},
month={Dec},
day={03},
volume={19},
number={1},
pages={31},
issn={1573-1332},
doi={10.1007/s11128-019-2527-8},
url={https://doi.org/10.1007/s11128-019-2527-8}
}

@Article{KRI2025,
author={Krishna, A. S. Abay
and Naseeda, K. K.
and Randeep, N. C.},
title={Bidirectional quantum teleportation using quantum walks},
journal={Quantum Information Processing},
year={2025},
month={Oct},
day={24},
volume={24},
number={11},
pages={348},
issn={1573-1332},
doi={10.1007/s11128-025-04965-6},
url={https://doi.org/10.1007/s11128-025-04965-6}
}

@Article{CHO2024,
author={Choudhury, Binayak S.
and Mandal, Manoj Kumar
and Samanta, Soumen},
title={Remote State Preparation of qubits Using Quantum Walks in the Presence of Controller},
journal={International Journal of Theoretical Physics},
year={2024},
month={Mar},
day={11},
volume={63},
number={3},
pages={71},
issn={1572-9575},
doi={10.1007/s10773-024-05584-0},
url={https://doi.org/10.1007/s10773-024-05584-0}
}
\end{document}